\documentclass[journal=cmatex,manuscript=article]{achemso}

\usepackage[version=3]{mhchem}
\usepackage{graphicx}
\usepackage{multirow}
\usepackage{color}
\usepackage{soul}
\usepackage{amsmath,amssymb}
\usepackage[hidelinks]{hyperref}

\newcommand{\llzo}{Li$_7$La$_3$Zr$_2$O$_{12}$}

\setkeys{acs}{articletitle = true}
\SectionNumbersOff
\AbstractOn

\title{Particle Morphology and Lithium Segregation to Surfaces of the Li$_7$La$_3$Zr$_2$O$_{12}$ Solid Electrolyte}

\author{Pieremanuele Canepa} \email{p.canepa@bath.ac.uk}
\affiliation{Department of Chemistry, University of Bath, Bath, BA2 7AY, UK}

\author{James A. Dawson} 
\affiliation{Department of Chemistry, University of Bath, Bath, BA2 7AY, UK}

\author{Gopalakrishnan Sai Gautam}
\affiliation{Department of Mechanical and Aerospace Engineering,
Princeton University, Princeton, NJ 08544, USA}

\author{Joel M. Statham}
\affiliation{Department of Chemistry, University of Bath, Bath, BA2 7AY, UK}

\author{Stephen C. Parker}
\affiliation{Department of Chemistry, University of Bath, Bath, BA2 7AY, UK}

\author{M. Saiful Islam} \email{m.s.islam@bath.ac.uk}
\affiliation{Department of Chemistry, University of Bath, Bath, BA2 7AY, UK}

\begin{document}

\begin{abstract}
Solid electrolytes for solid-state Li-ion batteries are stimulating considerable interest for next-generation energy storage applications. The \llzo\ garnet-type solid electrolyte has received appreciable attention as a result of its high ionic conductivity. However, several challenges for the successful application  of solid-state devices based on \llzo\ remain, such as  dendrite formation and maintaining physical contact at interfaces over many Li intercalation/extraction cycles. 
Here, we apply first-principles density functional theory  to  provide insights into the  \llzo\ particle morphology under various physical and chemical conditions. 
Our findings indicate Li segregation at the surfaces, suggesting Li-rich grain boundaries at  typical synthesis and sintering conditions. On the basis of our results, we propose practical strategies to curb Li segregation at the \llzo\ interfaces.   This approach can be extended to other Li-ion conductors for the design of practical energy storage devices.
\end{abstract}

\section{Introduction}
\label{sec:intro}
The commercial Li-ion battery, which relies on liquid electrolytes, is now the workhorse behind the mobile electronics industry.\cite{Armand2008,Dunn2011,Islam2014,Nykvist2015,Janek2016}
Unfortunately, a practical limit of what can be achieved with the current Li-ion technology is encountered when the focus shifts to electric vehicles.\cite{Dunn2011,Nykvist2015,Janek2016,Canepa2017} One promising avenue to improve the energy and power densities of Li-ion batteries, while enhancing their safety, consists of replacing the flammable liquid electrolyte with a solid electrolyte capable of efficiently shuttling Li ions between electrodes.\cite{Nb2003,Murugan2007,Knauth2009,Shinawi2013,Thangadurai2014,Kamaya2011,Masquelier2011,Wang2015,Bachman2016,Kato2016,Deng2015a,Mukhopadhyay2015,Deng2017,Lotsch2017,Kim2017}  

To facilitate this transition, the Li-ion conductivity of solid electrolytes must be competitive to that of their liquid analogs.\cite{Kamaya2011,Kato2016}  While significant attention is still devoted to intrinsic Li$^{+}$ conductivity in solid electrolytes, many challenges remain for  future solid-state applications.\cite{Cheng2013,Buschmann2011,Luntz2015,Richards2016,Janek2016,Yu2016b,Kerman2017,Yonemoto2017,Porz2017,Hanft2017} The most pressing challenges are finding solid electrolytes that are electrochemically stable against electrodes,  maintaining physical contact between components over many Li intercalation/extraction cycles and suppressing Li-dendrite formation. 

The \llzo\ garnet-type electrolyte has received significant attention due to its high ionic conductivity (10$^{-6}$--10$^{-3}$~S~cm$^{-1}$) achieved by a variety of doping strategies,\cite{Nb2003,Murugan2007,Geiger2011,Kuhn2011,Allen2012,Adams2012a,Shinawi2013,Thangadurai2014,Sharafi2017,Yi2017} but most importantly because of its perceived stability against the Li-metal anode.\cite{Cheng2015a,Richards2016,Ma2016,Han2016,Kim2016,Kerman2017,Porz2017,Wang2017} 

However, the failure of polycrystalline \llzo\ in  solid-state battery prototypes comprised of Li-metal anodes has been the subject of several studies.\cite{Kotobuki2010,Ma2016,Richards2016,Kim2016,Sharafi2016,Wang2017,Yonemoto2017}  It has been observed\cite{Porz2017} that once  Li fills a crack in Li$_6$La$_3$ZrTaO$_{12}$, fresh electro-deposited Li is extruded to the available surface.  Tests with  Li-metal/\llzo/Li-metal cells showed that only small current densities of $\sim$~0.5~mA~cm$^{-2}$ could be tolerated before dendrite failure.\cite{Kerman2017,Wang2017a} Rationalising the mechanisms behind the propagation of dendrites in \llzo\ is  a major challenge.  

In parallel, sintering strategies to maximise the bulk transport in ceramic materials are routinely applied. While high temperature densification enhances ion transport, the extent of morphological transformations of the electrolyte particles is still unclear. Kerman~\emph{et al.}\cite{Kerman2017} highlighted the connection between the processing conditions of  \llzo\ and its  particle morphology and size.  Kingon \emph{et al.}\cite{Kingon1983} demonstrated that ceramics containing volatile cations, such as \llzo, become Li deficient upon sintering. 

These experimental observations indicate that it is crucial to understand the variation of the \llzo\  morphology as a function of chemical and physical properties (composition and temperature). 

In this study, we develop a phenomenological model based on first-principles calculations to determine the  composition of \llzo\ particles, while the chemical environment of Li, La, Zr and O, voltage and/or the temperature are varied. Rationalising the particle morphology of solid electrolytes contributes towards a deeper understanding of several critical phenomena, including the Li$^+$ conductivity at  grain boundaries and the  propagation of dendrites during battery operation. Indeed, our results predict significant Li accumulation at the exterior of  the \llzo\ particles when we mimic reducing high-temperature synthesis conditions. 

Based on our computational insights, we propose practical strategies to engineer the chemical compositions of the particles, providing a greater control of the complex chemistry of  \llzo. These general design strategies  can be extended to other  solid electrolytes and electrode materials. 

\section{Results}
\label{sec:results}
\subsection{Phase stability and chemical domains}
\label{subsec:stablity}
\begin{figure*}[ht]
\includegraphics[scale=0.5]{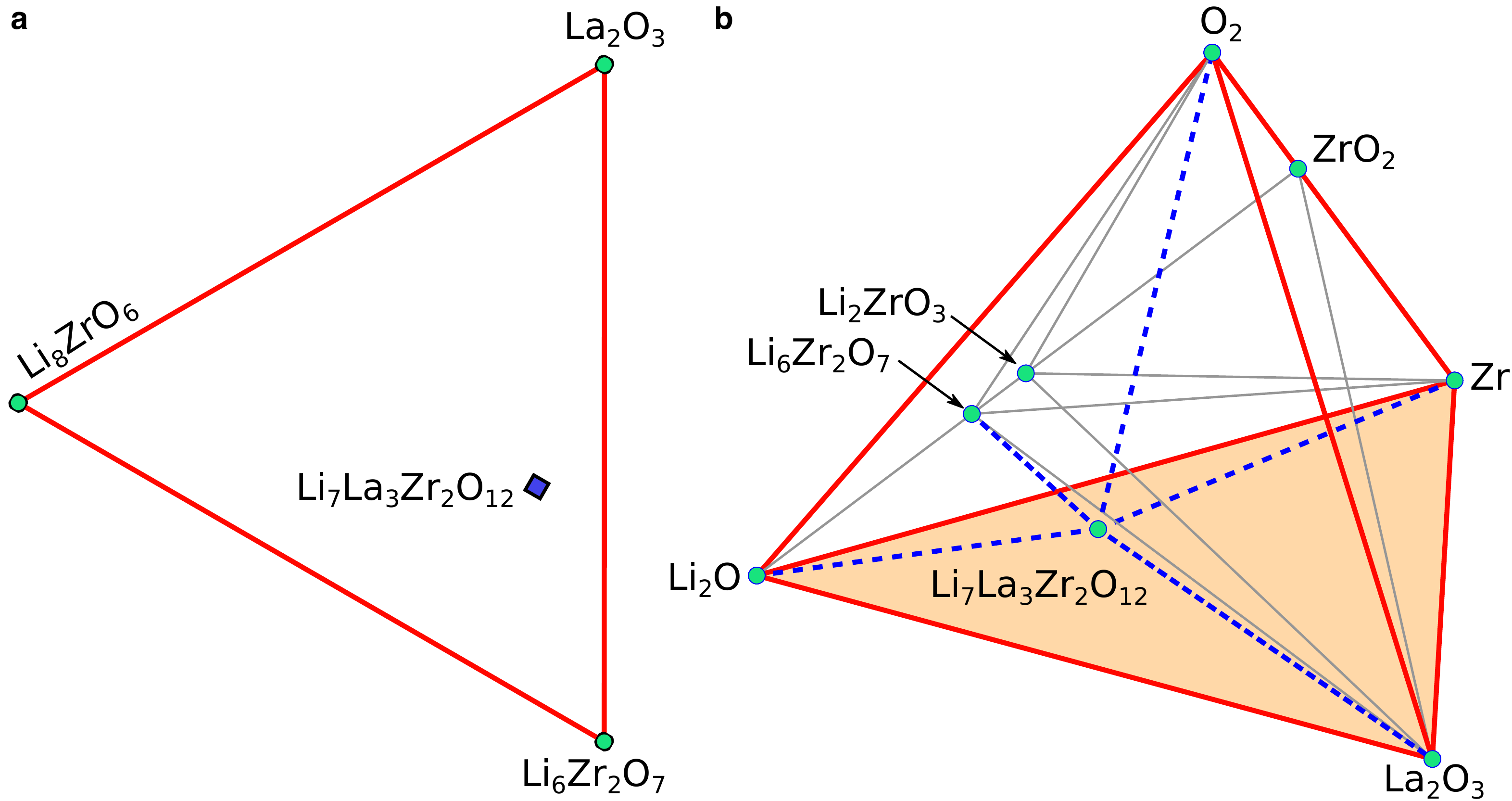} 
\caption{
({\bf a}) La$_2$O$_3$--Li$_6$Zr$_2$O$_7$-Li$_8$ZrO$_6$ projection of the quaternary Li--La--O--Zr phase diagram showing the decomposition products of the metastable \llzo\ (blue diamond), which are  Li$_6$Zr$_2$O$_7$, La$_2$O$_3$ and Li$_8$ZrO$_6$. ({\bf b}) Compound Li$_2$O--La$_2$O$_3$--Zr--O$_2$ phase diagrams where \llzo\ is assumed to be stable. Green dots display the stable phases, while red, blue and grey lines identify equilibrium tie lines. Dash blue lines mark tie lines shared by \llzo\ and some of the binary precursors used in its synthesis. Both phase diagrams are computed from DFT data at 0~K and combined with existing entries in the Materials Project database.\cite{Jain2013}
\label{fig:pd}
}
\end{figure*}

We first consider the relative stability of the \llzo\ tetragonal (space group $I$4$_1/acd$)  and high-temperature cubic ($Ia3\bar{d}$)  polymorphs.  The computed lattice constants ($a$ = 13.204 and $c$ = 12.704 \AA{}) of the tetragonal phase compare well with the experimental data ($a$ = 13.134 and $c$ = 12.663 \AA{}).\cite{Awaka2009}  Figure~\ref{fig:pd}a shows the decomposition of \llzo\ into Li$_6$Zr$_2$O$_7$ + La$_2$O$_3$ + Li$_8$ZrO$_6$, revealing the metastability of both the cubic ($\sim$~22~meV/atom above the stability line at 0~K) and tetragonal ($\sim$~7~meV/atom) polymorphs,  in agreement with previous density functional theory (DFT) preditions.\cite{Miara2013,Thompson2017}

The degree of metastability of the tetragonal phase is small enough that the compound can be stabilised by thermal effects, which explains the success of high-temperature ($>$ 600 $^{\circ}$C) phase-pure synthesis.\cite{Murugan2007,Thangadurai2014} It is assumed that the chemical decomposition of \llzo\ into Li$_6$Zr$_2$O$_7$ + La$_2$O$_3$ + Li$_8$ZrO$_6$ requires a major coordination rearrangement of Zr and La, thereby kinetically preventing Li$_7$La$_3$Zr$_2$O$_{12}$ from decomposing.   Therefore, we assume that \llzo\ is  thermodynamically stable (see phase diagram in Figure~\ref{fig:pd}b). 

Identifying the phases in equilibrium with \llzo\ (Figure~\ref{fig:pd}b) allows us to set the bounds of chemical potentials of each element, thus providing a thermodynamic framework to calculate meaningful non-stoichiometric surface energies (see Method section). Figure~\ref{fig:pd}b illustrates the phases in equilibrium with \llzo, which show that only La$_2$O$_3$, Li$_6$Zr$_2$O$_7$, Li$_2$O, O$_2$ and Zr are in direct equilibrium with the solid electrolyte.    Experimentally, the binary compounds La$_2$O$_3$, Li$_2$CO$_3$ (LiOH or Li$_2$O)\cite{Porz2017} and ZrO$_2$ are used as precursors for the synthesis of \llzo.\cite{Murugan2007,Porz2017} In addition,  when {\llzo} is assumed to be stable, Li$_8$ZrO$_6$ becomes metastable in the Li-La-Zr-O phase diagram (Figure~{\ref{fig:pd}}b).

Our discussion moves to the definition of the relevant chemical potentials, which have to be rigorously defined to accurately calculate the energies of non-stoichiometric surface structures (see Eq.~\ref{eq:gamma}).  From thermodynamic arguments,  any combination of three compounds in equilibrium with \llzo\  define  distinct chemical potentials ($\mu$) for the elements O, La, Li and Zr.  In this study, we consider two different chemical regimes, i.e. \emph{oxidising} and \emph{reducing}. The tetrahedron composed of \llzo, La$_2$O$_3$, Li$_2$O  and O$_2$ mimics the \emph{oxidising} and experimental synthesis conditions of \llzo.  In contrast, we consider a \emph{reducing} environment as defined by {\llzo}~being in equilibrium with Zr metal, La$_2$O$_3$ and Li$_2$O, which corresponds to experimental sintering conditions. A detailed derivation and the bounds of the chemical potential  used for each species are summarised in Section 1 and Table S1 of the Supplementary Information (SI). 

Although Zr forms oxides with multiple oxidation states, such as ZrO and Zr$_2$O as reported by Chen \emph{et al.},{\cite{Chen2015a}} Zr is not redox active in {\llzo}. Therefore, Zr-metal and ZrO$_2$ represent valid reference states for the $\mu_ {\rm Zr}$ in reducing (Zr$^0$) and oxidising (Zr$^{4+}$) environments, respectively.

\subsection{Surface structures and energies}
Surfaces of solid electrolytes are important to their electrochemical properties, particularly due to the presence of active interfaces within intercalation batteries.
The \llzo\ cubic polymorph provides the highest ionic conductivity.\cite{Thangadurai2014,Miara2013,Burbano2016} However, accounting  for the Li disorder presents a major computational complexity when creating representative surface structures. Thus,  we  consider the tetragonal polymorph, which constitutes a distinct ordering of Li sites, as the reference structure for creating our surface models. 

\begin{figure}[!ht]
\includegraphics[width=\columnwidth]{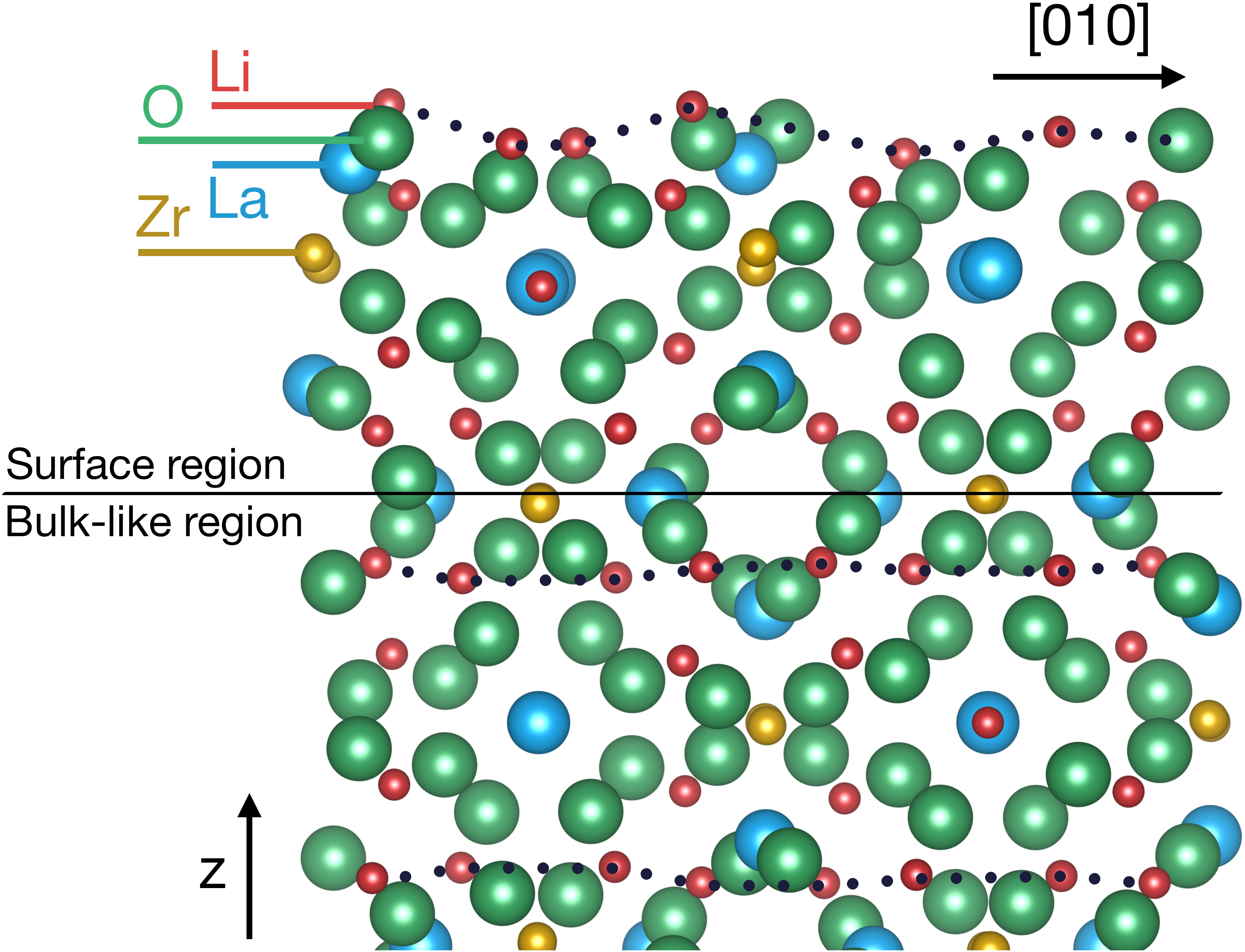}
\caption{ Sideview of the non-stoichiometric (010) Li-terminated surface of  \llzo. Li atoms  are in red, O in green, La in blue and Zr in gold. Solid lines identify the arrangement of each atom plane along the non-periodic $z$ axis. The black line marks the separation between the bulk-like region from the surface region. The  black dotted lines are  guides for the eye to highlight the change in the local Li symmetry upon surface reconstruction.
\label{fig:stru}
}
\end{figure}

Figure~\ref{fig:stru} depicts the atomic arrangement of the Li-terminated (010) surface of \llzo, highlighting the significant reconstruction of the Li and O layers, respectively. 
The dotted lines in Figure~\ref{fig:stru} are a guide for the eye to illustrate the loss of symmetry of the Li environment at the surface compared to the bulk region.  Figure~{\ref{fig:stru}}  shows that La layers overlap with ``rumpled'' oxygen layers, which contribute to stabilise La-terminated surfaces, as discussed in the following paragraphs. In the case of Zr ions, the oxygen coordination environment in the surface slab show insignificant deviation from the octahedral coordination within bulk {\llzo}, in qualitative agreement with the lack of surface reconstruction observed in ZrO$_2$.{\cite{Haase1998}}

\begin{figure}[!ht]
\includegraphics[scale=0.5]{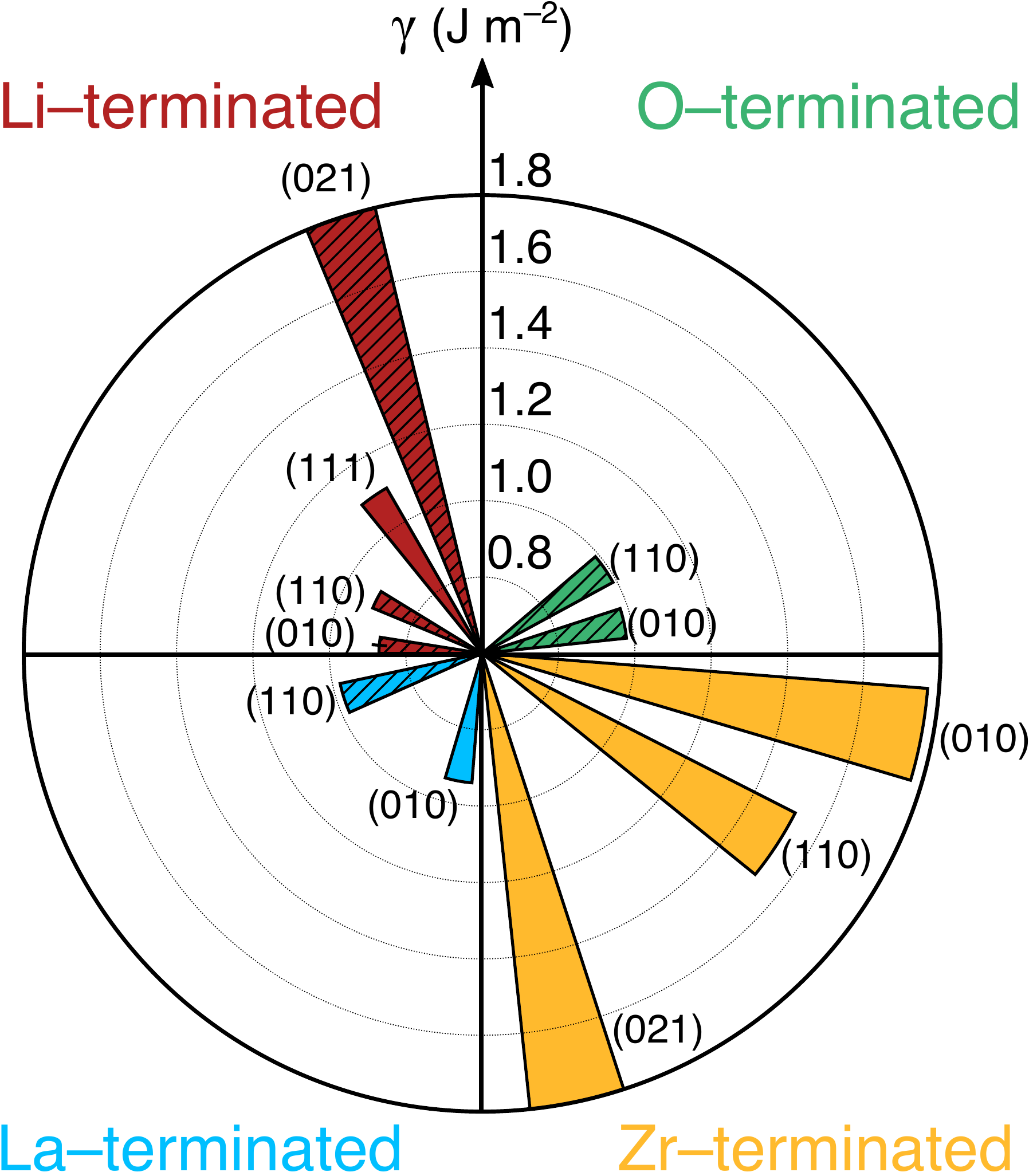} 
\caption{Surface energies $\gamma$ (J m$^{-2}$) of La (blue), Li (red), O (green) and Zr (yellow) -terminated surfaces of \llzo.  Hatched bars indicate non-stoichiometric surfaces, whose surface energies are derived using the chemical potentials from Figure~\ref{fig:pd}b.   The chemical potentials of Li, La and Zr are fixed by Li$_2$O,  La$_2$O$_3$ and Zr metal, respectively, corresponding to reducing conditions (details in Section~{\ref{subsec:stablity}}).
\label{fig:gammas}
}
\end{figure}

It is known that for a given Miller index several surface terminations may be possible since the bulk can be cleaved at different planes, as shown in Figure~\ref{fig:stru}.
The relative stability of each surface model is defined by their surface energy ($\gamma$, Eq.~\ref{eq:gamma}).
Figure~\ref{fig:gammas} depicts the computed $\gamma$ values of a number of stoichiometric and non-stoichiometric La, Li, O and Zr-terminated surfaces of \llzo.  Non-stoichiometric surfaces refer to surfaces where the stoichiometry deviates in composition from the bulk. The surface energies of symmetry-related Miller index surfaces (e.g., (100) $\approx$ (010) $\approx$ (001)) are detailed in Table S2. As introduced in Section~\ref{subsec:stablity}, the chemical potentials, $\mu_i$, for calculating $\gamma$ of non-stoichiometric surfaces are set to reducing conditions (i.e., $\mu_{\rm La} \approx \mu_{\rm La}$ in La$_2$O$_3$, $\mu_{\rm Li}  \approx \mu_{\rm Li}$ in Li$_2$O, and $\mu_{\rm Zr}  \approx \mu_{\rm Zr}$ in Zr metal), see Figure~{\ref{fig:pd}}b and Section~1 of the SI.

Figure~\ref{fig:gammas} shows three main features:\  i) Zr-terminated surfaces show the highest surface energy $\gamma$ ($>$~1.5~J~m$^{-2}$), ii) certain Li-terminated surfaces possess significantly lower $\gamma$ ($\sim$~0.87$\pm$0.02~J~m$^{-2}$ for the (010) surface),  in good agreement with previous work.\cite{Thompson2017} iii) La- and O-terminated surfaces show similar surface energies, as indicated by $\gamma$~0.98~J~m$^{-2}$ and 0.99~J~m$^{-2}$ for the La- and O-terminated (110) surfaces, respectively. 

Although the surface structures are obtained from the tetragonal phase, we find identical surface energies for symmetry inequivalent surfaces (see Table S1 and Figure S1). For example, the surface energy ($\sim$~1.77~J~m$^{-2}$) of the Zr-terminated (010) surface is identical to the (001) and (100) surfaces, which is typically not found for tetragonal  structures. This suggests the similarity between the tetragonal and cubic phases of \llzo ~and indicates that the Li ordering, which affects the relative stability of the bulk tetragonal and cubic phases, has only a negligible impact on the relative symmetry and energetics of {\llzo}~surfaces. Notably, the $c/a$ ratio exhibited by the tetragonal phase ($\sim 0.96$ from experimental lattice constants, see Section~{\ref{subsec:stablity}}) signifies the ``small'' tetragonal distortion in {\llzo}.

\subsection{Effects of oxygen environment and temperature on surfaces}
\noindent With the aim of understanding the interplay between compositional and temperature effects on the morphology of \llzo, we now move our attention to trends of surface energy as a function of temperature and oxygen composition.   To include  temperature dependence in our model, we apply a thermodynamic framework (detailed in the Method section) that connects changes in the O$_2$ chemical potential, $\mu_{\rm O_2}$, directly to temperature.\cite{Reuter2001} This approximation is valid as the \llzo\ electrolyte is in contact with an oxygen environment during its synthesis and sintering.  

With $\mu_{\rm O}$  = $\frac{1}{2} \mu_{\rm O_2}$, $\mu_{\rm O}$  sets the surface energy of non-stoichiometric surfaces, as indicated in Eq.~{\ref{eq:gammaT}}. Note that under both oxidising and reducing conditions (Section~{\ref{subsec:stablity}}), the chemical potentials of La and Li are set by La$_2$O$_3$ and Li$_2$O, respectively.  All the non-stoichiometric surfaces studied here are either oxygen rich or poor (see Method section).  High $\mu_{\rm O}$ (or $\mu_{\rm O_2}$) represents low-temperature situations and a highly oxidative environment, where oxygen molecules ``condense'' on the surfaces of \llzo. In contrast, higher temperatures (i.e., more negative $\mu_{\rm O}$) signify reducing conditions, where oxygen atoms become volatile and leave the surface as O$_2$ gas, which is equivalent to {\llzo}~being in equilibrium with Zr metal.

\begin{figure}[!h]
\includegraphics[width=\columnwidth]{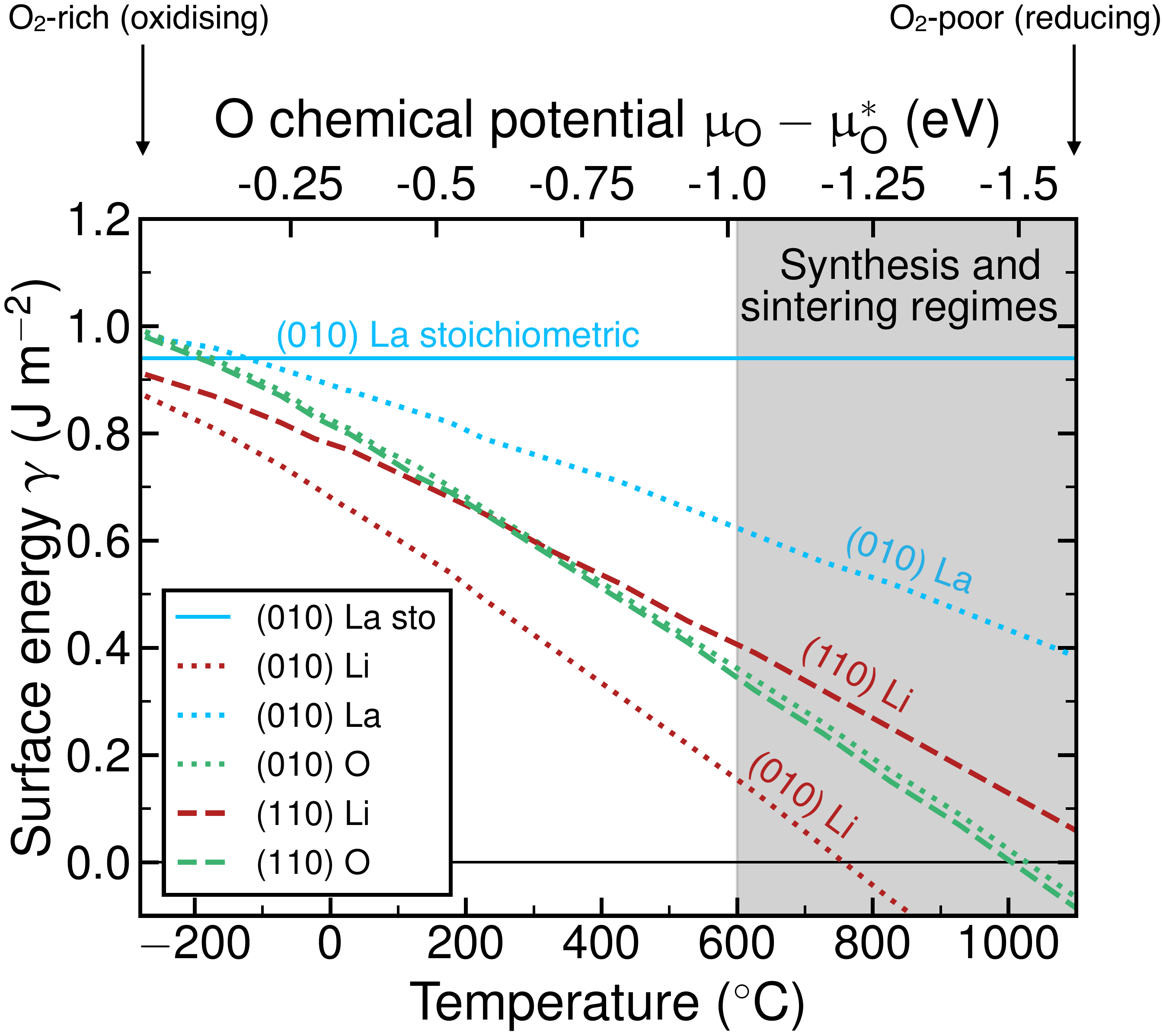}
\caption{Surface energy $\gamma$ of La (blue), Li (red) and O (green) -terminated \llzo\ surfaces vs temperature and oxygen chemical potential $\mu_{\rm O}$. The  blue horizontal line indicates the stoichiometric La-terminated surface energy. The zero (eV) in the $\mu_{\rm O}$ scale is normalised against the reference state $\mu_{\rm O}  ^*$ and is  detailed in the SI. $\mu_{\rm O}$ near 0~eV relates to oxygen-rich (or oxidising) regimes, whereas more negative oxygen chemical potentials are oxygen-poor (or reducing) conditions. The grey shading marks the experimental temperature window for synthesis and sintering of \llzo.\cite{Thangadurai2014}  The chemical potentials of Li and La are fixed by Li$_2$O and La$_2$O$_3$, respectively, while  $\mu_{\rm O}$ is allowed to vary.  
\label{fig:gammat}
}
\end{figure}

Figure~\ref{fig:gammat} shows the variation of the surface energy  for a number of non-stoichiometric surfaces as a function of temperature, or its equivalent  $\mu_{\rm O}$, corresponding to an oxygen partial pressure of 1~atm.

A number of important observations can be drawn from Figure~\ref{fig:gammat}. i) The Li-terminated (010) surface has the lowest $\gamma$ (as in Figure~\ref{fig:gammas})  and the La-terminated (010) stoichiometric surface has the highest $\gamma$ for temperatures higher than 25~$^{\circ}$C.  The negative slope of each line signifies that all the surfaces are oxygen deficient. While studying non-stoichiometric surfaces, we have focused on La, Li and O deficient scenarios, as they are most likely to develop at high temperatures.\cite{Antoini1992,Li_2013} In order to maintain the electroneutrality of oxygen-terminated surfaces, oxygen vacancies were introduced to compensate the removal of cations  (details are provided in the Method section). ii) The  stability of the Li-terminated (010) surfaces in comparison to other terminations is significant. iii) At temperatures higher than 300~$^{\circ}$C, the O-terminated (010) and (110) surfaces become more stable than the (110) Li-terminated surface. This result is also found for the (100), (001), (011) and (101) Li-terminated facets. iv) Above 750~$^{\circ}$C, the negative $\gamma$ of (010) Li-terminated surface (as seen in Figure~\ref{fig:gammat}) is indicative of the instability of bulk {\llzo}, and may be linked to the  melting of {\llzo} particles. 

\subsection{Environment dependent particle morphologies}
By combining our surface energies of various surface facets at distinct chemical compositions (Figures~\ref{fig:gammas} and~\ref{fig:gammat}), we can implement the Wulff construction to derive the  \llzo\  equilibrium particle morphology at synthesis conditions.

\begin{figure}[!th]
\includegraphics[width=\columnwidth]{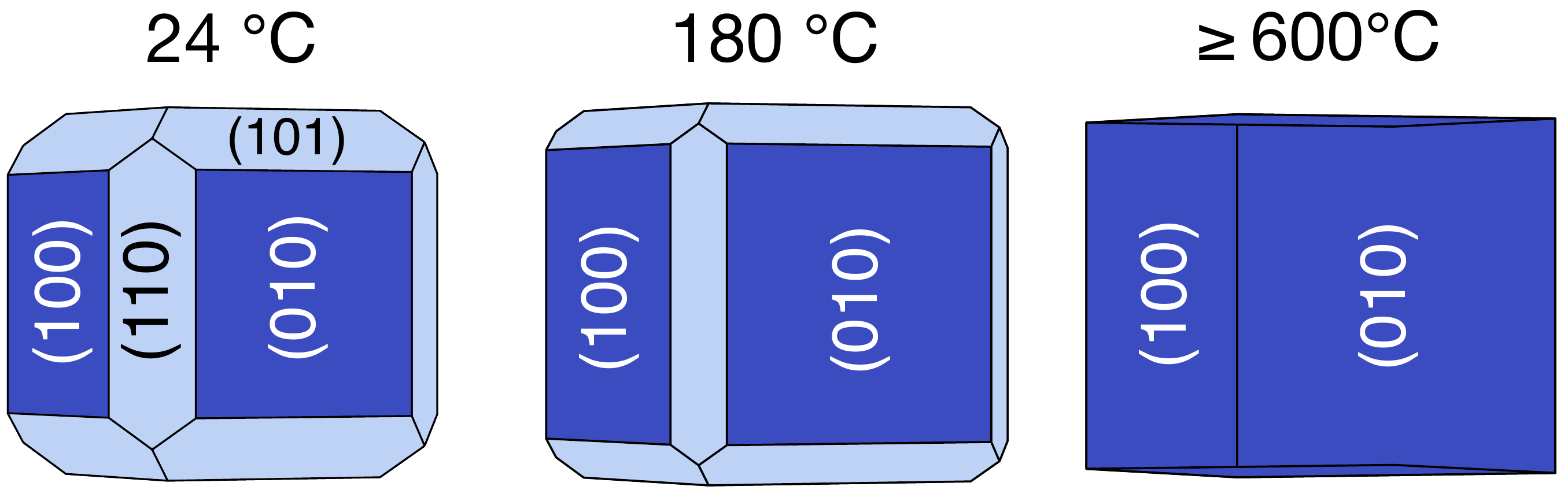} 
\caption{Variation in the \llzo\ equilibrium morphology with increasing temperature. The particles are expected to be Li-terminated, as suggested in Figure~\ref{fig:gammat}. Labels identify the surface planes of interest. 
\label{fig:wulff}
}
\end{figure}

Figure~\ref{fig:wulff} depicts the change of the particle equilibrium morphology as a function of temperature.

 At room temperature ($\sim$~24 $^{\circ}$C), the equilibrium {\llzo} particle morphology is dominated by the (001), (101) and (110) surfaces. For temperatures greater than 600~$^{\circ}$C (and $< 750~^{\circ}$C), {\llzo} particles should assume a cubic shape dominated by the (100) and (010) surfaces, as seen in Figure~\ref{fig:wulff}.

At 24~$^{\circ}$C and intermediate temperatures ($\sim$~180~$^{\circ}$C), the (110)  Li-terminated surface contributes to the overall particle shape. However,  an increase in oxygen composition on the surface of the {\llzo} particles will be also observed, as shown by the increased stability of the (110) oxygen-terminated surfaces over (110) Li-terminations, as seen in Figure~\ref{fig:gammat} at temperatures above 300 $^{\circ}$C.

\subsection{Tuning the synthesis conditions of \llzo}
We now discuss the surface phase diagram obtained by varying the chemical composition of \llzo. This analysis contributes to understanding the experimental synthesis conditions to achieve the desired chemical composition of the particle surfaces. 

Computing a complete surface phase diagram represents a formidable exercise given the large compositional space for the non-stoichiometric terminations accompanied by the large number of  atomic arrangements of partially occupied terminating layers. Thus, we limit the discussion of the surface phase diagram to the \llzo\ surfaces in Figure~\ref{fig:gammas}. Li-rich and Li-poor conditions correspond to Li$_2$O (reducing conditions) and Li$_6$Zr$_2$O$_7$ (oxidising conditions), respectively, while Zr-rich is equivalent to Zr metal (reducing) and Zr-poor to O$_2$ gas (oxidising). 

\begin{figure*}[h!t]
\includegraphics[scale=0.45]{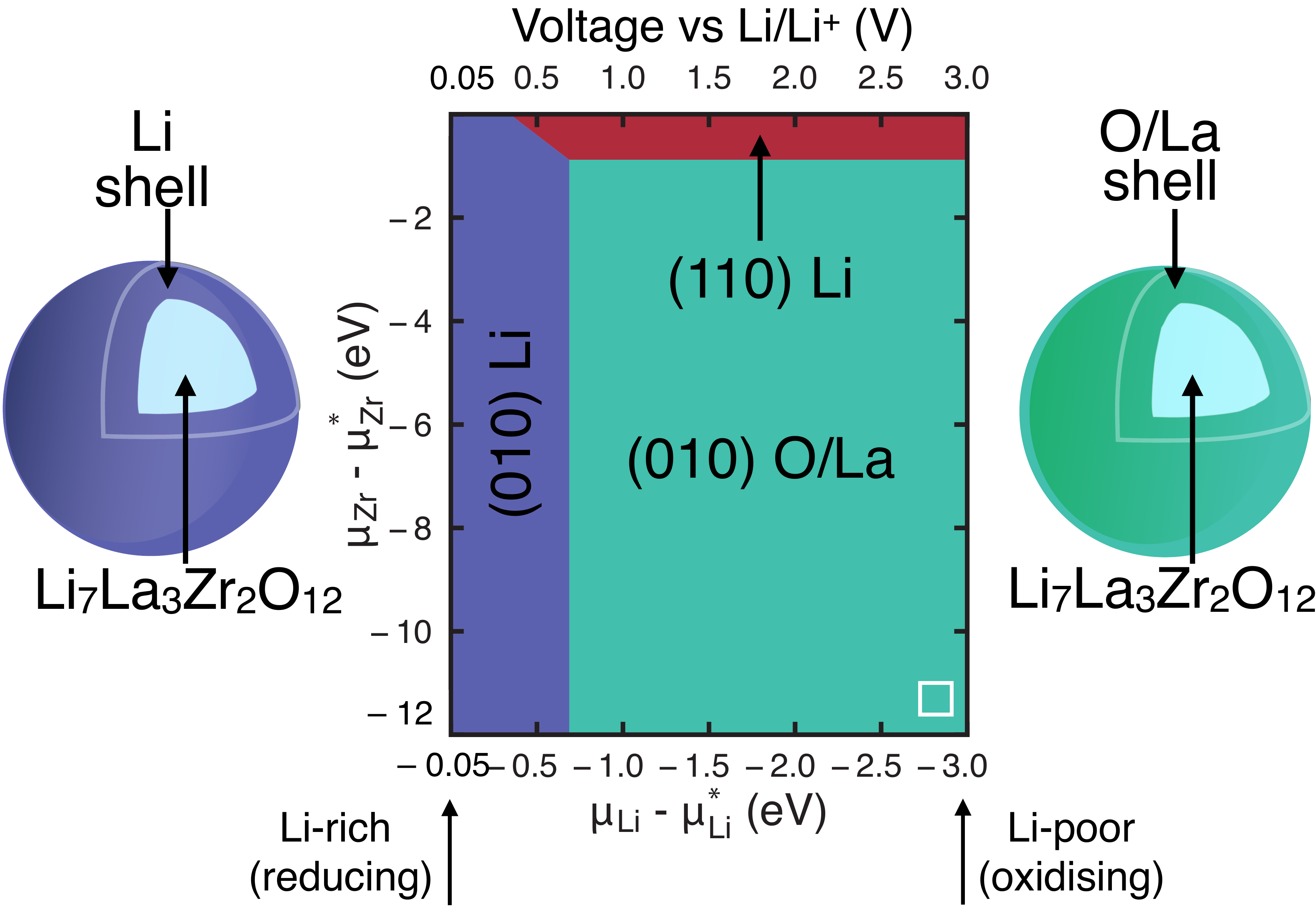} 
\caption{Surface phase diagram at 0 K of \llzo\ and schematic representations of particle morphologies at different chemical conditions. Stable surfaces and chemical terminations as a function of $\mu_{\rm Li}$ and $\mu_{\rm Zr}$. The white square identifies the compositional Li--Zr conditions where \llzo\ is commonly synthesised. Zr-rich is equivalent to Zr metal (Zr-poor is O$_2$ gas), whereas Li-rich is Li metal (and Li-poor is Li$_6$Zr$_2$O$_7$).  The voltage evolution vs Li/Li$^*$ (with ${\rm V} = - \mu_{\rm Li}\cdot e^-$) is also shown. The chemical potential scales are referenced against the reference states $\mu_{\rm Li}  ^*$ (Li$_2$O) and $\mu_{\rm Zr}  ^*$ (Zr metal). 
\label{fig:pdS}
}
\end{figure*}

Figure~\ref{fig:pdS} shows the surface phase diagram at 0 K by varying the Li ($\mu_{\rm Li}$) and Zr ($\mu_{\rm Zr}$) composition. 
We find that regions of low $\mu_{\rm Li}$ and $\mu_{\rm Zr}$ are consistently dominated by the  (010)  O- and La-terminated surfaces.  At more positive $\mu_{\rm Li}$ and $\mu_{\rm Zr}$ (near Li-rich and Zr-rich conditions), the (010) Li-terminated surfaces are stable.   In fact,  the  (010) O- and La-terminated surfaces have similar surface energies $\sim$~0.94 and  $\sim$~0.98 J~m$^{-2}$, respectively (see Figure~\ref{fig:gammas}), as the La ions exposed are surrounded by a O sub-layer.  The La/O or Li segregations at the surface of the particles of \llzo\  (at specific $\mu_{\rm Li}$ and $\mu_{\rm Zr}$) are schematically shown by the green and violet spheres of Figure~{\ref{fig:pdS}}.

Figure~\ref{fig:pdS} also includes a voltage scale, which relates directly to the Li chemical potential (${\rm V} = - \mu_{\rm Li}\cdot e^-$).  Negative $\mu_{\rm Li}$ signify high voltages (vs Li/Li$^{+}$) and vice versa. 

\section{Discussion}
\label{sec:discussion}
To gain realistic insights into the design of solid electrolytes for solid-state batteries, we have performed a thorough first-principles calculation analysis of the \llzo\ surfaces and its morphologies under various physical and chemical conditions. 

\noindent  {\bf Morphology and composition of \llzo\ particles --} Figure~\ref{fig:pd}b shows that lower surface energies are found for surfaces terminated by cations with lower oxidation states, following the trend Li$^+ <$ La$^{3+} <$ Zr$^{4+}$. This finding  relates to electrostatic and geometric factors. By cleaving a cation-terminated surface, the large disruption of the ideal cation coordination environment results in a high-energy penalty, thus impacting significantly the relative stability of the surface.  

Experimentally,\cite{Brown1988} it is found that La$^{3+}$ and Zr$^{4+}$ ions prefer high oxygen coordination ($\geq$~6 in the cubic and tetragonal \llzo\ phases), whereas Li$^+$ can  adjust to both octahedral and tetrahedral environments.\cite{Thangadurai2014,Rong2015}  Li ions can tolerate reduced coordination environments leading to lower surface energies  compared to Zr-terminated surfaces, which undergo a reduction in coordination from 6--8 to 4.  La-terminated surfaces show low surface energies ($\sim$~0.94~J~m$^{-2}$) compared to the Zr-terminated surface, which are explained by the oxygen sub-layer stabilising the partially uncoordinated La atoms and lowering the  surface energy (see Figure~S3). 

We have identified that surfaces with low Miller indices, e.g.\ (010) and (110) with Li-rich textures, dominate across a wide range of temperatures and oxygen environments. Li segregation at the surfaces of  \llzo\ particles has been demonstrated by neutron depth profiling experiments.\cite{Han2016}  O-terminated surfaces are also possible, as shown in Figure~\ref{fig:pdS}. This may be significant in relation to the recent report of oxygen migration in \llzo.\cite{Kubicek2017}

The predicted room temperature morphology of \llzo\  is in excellent agreement with a scanning electron microscopy study of a single crystal,\cite{Awaka2009} providing credibility to the computed morphologies of Figure~\ref{fig:wulff}. However, no specific surface facets were characterised, which we identified here.\cite{Awaka2009} We can complement the experimental observations by extending our model beyond the shape of the particles. This is completed by ascertaining the dominant surface facets and the most likely chemical compositions under both reducing and oxidising conditions. 

On the basis of these findings we speculate that small cations, such as Al$^{3+}$  and Ga$^{3+}$, doped at Li$^{+}$  sites may segregate at the surfaces of the particles. In agreement with our hypothesis, a number of experimental reports demonstrate that  Al$^{3+}$ segregates to the grain boundaries of doped \llzo.\cite{Li_2013,Cheng2014,Ohta2016} We speculate that high-valent cations, such as Ta$^{5+}$ and Bi$^{5+}$  (introduced on the Zr lattice to increase the number of Li vacancies),\cite{Thangadurai2014} will constitute the core of the \llzo\ particles.   \\

\noindent {\bf Densification and implications on ionic conductivity --} Densification of ceramic oxides via high-temperature (and spark-plasma) sintering is routinely employed to improve the electrolyte ionic conductivity.\cite{Kingon1983,Li_2013,Thangadurai2014}   Typically, the interpretation of impedance measurements  requires the deconvolution of the total ionic conductivity into three main contributions, namely, \cite{Jamnik1999,Jamnik2001,Lai2005} intrinsic bulk,  grain boundary and interfacial electrolyte/blocking electrode. While bulk Li-ion transport has been emphasised by both experiment and computation,\cite{Nb2003,Murugan2007,Knauth2009,Thangadurai2014,Geiger2011,Kuhn2011,Miara2013,Burbano2016} grain boundary Li-ion conductivity is much less examined, despite being crucial.\cite{Nb2003,Murugan2007,Knauth2009,Thangadurai2014} 

The seminal paper on \llzo\  by Murugan~\emph{et al.}\cite{Murugan2007} showed significant  Li-ion resistance at the grain boundaries ($\sim$~50\% of the total), thus suggesting the relevance of intergranular Li-ion transport.\cite{Dawson2017}   Ceramic oxides processed at high temperatures containing ``volatile'' cations, such as Li, including \llzo,  will  produce Li deficient bulk materials,{\cite{Kerman2017}}  and thus possible Li loss upon sintering treatments. For example, Antolini\cite{Antoini1992}  showed  that sintering of Li$_x$Ni$_{1-x}$O ceramic electrodes  can promote Li segregation at the particle exterior, thus altering the overall stoichiometry. Loss of Li$_2$O was observed in the synthesis of \llzo,\cite{Li_2013,Sharafi2017,Kazyak2017,Yi2017} and additional  Li$_2$O is routinely added during its preparation. 

In agreement, our prediction in Figure~\ref{fig:gammat}  suggests that at high temperatures ($\geq$~600~$^{\circ}$C in the sintering regime) and reducing conditions, the particle surfaces will show pronounced segregation of Li. Assuming that the stable surfaces computed in this study are representative of the grain boundaries in \llzo, we speculate that the accumulation of Li ions can impact the Li transport involving grain boundaries.\cite{Dawson2017} \\
 
\noindent {\bf Engineering the particle morphology --} On the basis of our predictions,  we can propose practical strategies to engineer particle morphologies of \llzo. 

For example, Figure~\ref{fig:pdS} demonstrates that adding extra  Zr and/or Li metals  during synthesis may promote  Li segregation at the grain boundaries.\cite{Han2016} In addition, as indicated in Figure~{\ref{fig:gammat}}, routine high-temperature synthesis of {\llzo} promotes reducing conditions (i.e., oxygen-poor conditions) and Li terminated surfaces/particles. Hence  low-temperature synthesis (and sintering) protocols should be sought.{\cite{Amores2016}}

From analysis of Figure~\ref{fig:pdS}, we speculate that O/La accumulation at the grains is also observed near the operating voltages of typical Li-ion cathode materials (e.g., LiCoO$_2$  $\sim$~3.8~V vs.\ Li/Li$^+$ and LiFePO$_4$ $\sim$~3.4~V). In this context, Miara \emph{et al.}\cite{Miara2015} have  shown that \llzo\ remains stable against LiCoO$_2$, whereas the analogous interface with LiFePO$_4$ decomposes forming a protecting Li$_3$PO$_4$ interface. Nevertheless, a more recent experimental investigation by Goodenough \emph{et al.},\cite{Park2016} showed significant Al$^{3+}$ and La$^{3+}$ migration from Al-doped \llzo\  to LiCoO$_2$, and negligible Zr  diffusion into LiCoO$_2$. In Figure~\ref{fig:pdS}, near 3~V we predict La segregation towards the particle surfaces  corroborating these experimental findings.

The failure upon short-circuiting of polycrystalline {\llzo} in solid-state devices, utilising a Li-metal anode, has been linked to dendrite propagation.{\cite{Kerman2017}
Near  0 V or at the potential of Li metal, we expect the \llzo\ particles to be lithium terminated. In agreement with our  results, Li segregation close to a Li-metal anode interface in Al-doped {\llzo}  has been recently observed by in situ transmission electron microscopy.{\cite{Ma2016}} We speculate that the occurrence of Li at the particle surface and at grain boundaries, could indeed set the ideal chemical environment required for Li-dendrite growth and propagation between \llzo\ particles.  In line with our results, Kerman \emph{et al.}\cite{Kerman2017} proposed that once Li fills a crack in doped \llzo, fresh electrodeposited Li extrudes to the existing grain boundaries. 

Unsurprisingly, the process of dendrite propagation can originate from Li ``stuffing" into grain boundaries.{\cite{Kerman2017}} Thus, the significant accumulation of Li at the surfaces of the {\llzo} particles may favour the initial stages of dendrite nucleation and  growth along the existing grain boundaries.}
Further experimental studies are required to verify these hypotheses.

\section{Conclusions}

 \llzo\ is an important solid electrolyte material, but its surfaces and particle morphologies under synthesis and sintering conditions are not fully characterised.

First, by studying the morphology and composition of \llzo\ particles from DFT-based calculations, we have demonstrated the spontaneous segregation of Li towards the particle exterior. Second, we map the compositional changes of the surfaces of Li$_7$La$_3$Zr$_2$O$_{12}$ as a function of temperature and of oxygen chemical pressure. Li segregation to surfaces is the dominant process over a range of temperatures, particularly during high-temperature synthesis and sintering. These findings are significant in relation to the initial stages of Li dendrite growth. Third, by studying the surface phase diagram of \llzo, we find that Li segregation  can be curbed by tuning the ceramic synthesis conditions.  We show that synthesis in reducing  environments (O-poor, Li-rich and/or Zr-rich) may  promote Li segregation to the particle surfaces. Finally, we find that particle compositions of \llzo\ are altered upon voltage sweeps, with Li segregation at the exterior occurring at the Li-metal anode voltage. 

These findings will contribute towards developing strategies for the optimisation of the synthesis and operation of  promising solid electrolytes for solid-state batteries.


\section{Method}
\label{sec:method}
\subsection{Surface energies and thermodynamic framework}
The physical quantity defining stable surface compositions and geometries is the surface free energy $\gamma$ (in J~m$^{-2}$):
\begin{equation}
\label{eq:gamma}
\gamma = \frac{1}{2A} \left[ G_{\rm surface} - G_{\rm bulk} - \sum _{i} ^{\rm species} \Delta n_i \mu_i \right]
\end{equation}
where $A$ is the surface area (in m$^{-2}$) and $G_ {\rm surface}$ and $G_ {\rm bulk}$ are the surface free energies of periodic surfaces and the reference bulk material, respectively. $G_ {\rm surface}$ and $G_ {\rm bulk}$  are approximated by the respective computed internal energies $E_ {\rm surface}$ and $E_ {\rm bulk}$, accessed by density functional theory (DFT) as described in the SI. In the case of non-stoichiometric surfaces, the final surface energy depends on the environment set by the chemical potential $\mu _{i}$ for species $i$ and amounting to an off-stoichiometry of $\Delta$$n_i$. Note that $\Delta n_i$ is negative (positive) if species $i$ is removed (added) to the surface. The chemical potential references $\mu _i$ were derived from the computed phase diagram (Figure~\ref{fig:pd}b) at 0 K. 

Eq.~\ref{eq:gamma} provides $\gamma$ values at 0 K that are not representative for the operating conditions of solid electrolytes and the synthesis and sintering conditions. For non-stoichiometric surfaces, the approximation chosen to introduce the temperature dependence in the $\gamma$ values is based on the fact that the surrounding O$_2$ atmosphere forms an ideal-gas-like reservoir, which is in equilibrium with \llzo. The effect of temperature is introduced into the definition of $\gamma$ as follows:
\begin{equation}
\label{eq:gammaT}
\gamma(T) = \frac{1}{2A} \left[ E_{\rm surface} - E_{\rm bulk} - \sum _{i} ^{\rm species -\{O\} } \Delta n_i \mu_i  - \Delta n_{\rm O} \mu_{\rm O}(T) \right]
\end{equation}
\noindent where $\mu_{\rm O}$ is now a temperature dependent quantity and evaluated directly by combining DFT data with experimental values tabulated by NIST/JANAF as:\cite{Chase1996}
 \begin{equation}
\label{eq:mufinal}
\mu _{\rm O}(T) = \frac{1}{2} \mu _{\rm O_2} ({\rm 0\, K, DFT}) + \frac{1}{2} \mu _{\rm O_2} ({\rm 0\,  K, Exp.}) +\frac{1}{2}\Delta G_{\rm O_2} (\Delta T, {\rm Exp.})
\end{equation}
 where the $\mu _{\rm O_2} ({\rm 0\, K, DFT})$ is the 0 K free energy of an isolated oxygen molecule evaluated with DFT, whereas  $\mu _{\rm O_2} ({\rm 0\, K, Exp.})$ is the 0 K experimental (tabulated) Gibbs energy for oxygen gas. $\Delta G_{\rm O_2} (\Delta T, {\rm Exp.})$ is the difference in the Gibbs energy defined at temperature, $T$, as $1/2[H(T, {\rm O_2}) - H(0\, K, {\rm O_2})] - 1/2T [S(T, {\rm O_2})]$, respectively, as available in the NIST/JANAF tables.\cite{Chase1996} We omitted the partial pressure dependence of the $\mu_{\rm O_2}$ term (i.e., we used $p_{\rm O_2}$~=~1~atm) as we expect this contribution to be small, as demonstrated previously.\cite{Reuter2001} 

 \subsection{Bulk surface models}
Because of the large number of possible chemical terminations, as a result of the quaternary nature of \llzo, the selection of surfaces investigated was only limited to low-index surfaces, such as (100), (001), (101), (111) and (201). We note that some of these surfaces are related by the intrinsic tetragonal symmetry. For example, (100) = (010), as verified by the surface energies in the Supplementary Information (see Table S2). 

In line with Tasker's classification of oxide surfaces,\cite{Tasker1979} only realistic type I surfaces were considered, which are characterised by zero charge and no electrical dipole moment. Nevertheless, these requirements are only satisfied by a limited number of stoichiometric Zr- or La-terminated  surfaces with high surface energies.

 Because our goal is to rationalise the chemical composition and morphology of the \llzo\ particles, it is crucial to study the Li- and O-terminated surfaces. As a result, type I non-stoichiometric surfaces were generated by selectively removing layers of Zr and/or La and charge-compensated by O removal, as shown schematically in Figure~\ref{fig:stru}. Upon cation removal, charge neutrally is maintained by introducing oxygen vacancies, resulting in the need to investigate a significant number of atomic orderings. We simplify this difficult task by computing with DFT only the 20 orderings with the lowest electrostatic energy, as obtained by minimising the Ewald energy of each surface using formal charges.\cite{Ong2013b} This results in the assessment of 420 non-stoichiometric surfaces using DFT. While performing this operation, we enforce symmetry between the two faces of the surfaces. Using this strategy, we identified 21 non-stoichiometric orderings and 11 stoichiometric surfaces that are O-, Li-, La- and Zr-terminated, respectively, whose surface energies are discussed in Figure~\ref{fig:gammas} and Table S2.



\section*{Acknowledgements}
The authors gratefully acknowledge the EPSRC Programme Grant (EP/M009521/1) and the MCC/Archer consortium (EP/L000202/1).
PC is grateful to the Ramsey Memorial Trust and University of Bath for the provision of his Ramsey Fellowship. PC is thankful to Dr.\ Benjamin Morgan at the University of Bath for fruitful discussions. PC deeply indebted to Dr. yet to be Theodosios Famprikis at the LRCS, Amiens, France. We acknowledge fruitful discussion with Prof.\ Peter Bruce, Stefanie Zekoll and Dr.\ Jitti Kasemchainan  at the University of Oxford. 

\begin{suppinfo}
The Supporting Information is available free of charge on the ACS Publications website at DOI: 
\begin{itemize}
\item Details of first-principles calculations.
\item Chemical potentials bounds.
\item Surface energies.
\item \llzo\ particle morphologies.
\item \llzo\ surface reconstruction.
\end{itemize}
\end{suppinfo}


\bibliography{biblio}

\providecommand{\latin}[1]{#1}
\makeatletter
\providecommand{\doi}
  {\begingroup\let\do\@makeother\dospecials
  \catcode`\{=1 \catcode`\}=2 \doi@aux}
\providecommand{\doi@aux}[1]{\endgroup\texttt{#1}}
\makeatother
\providecommand*\mcitethebibliography{\thebibliography}
\csname @ifundefined\endcsname{endmcitethebibliography}
  {\let\endmcitethebibliography\endthebibliography}{}
\begin{mcitethebibliography}{72}
\providecommand*\natexlab[1]{#1}
\providecommand*\mciteSetBstSublistMode[1]{}
\providecommand*\mciteSetBstMaxWidthForm[2]{}
\providecommand*\mciteBstWouldAddEndPuncttrue
  {\def\EndOfBibitem{\unskip.}}
\providecommand*\mciteBstWouldAddEndPunctfalse
  {\let\EndOfBibitem\relax}
\providecommand*\mciteSetBstMidEndSepPunct[3]{}
\providecommand*\mciteSetBstSublistLabelBeginEnd[3]{}
\providecommand*\EndOfBibitem{}
\mciteSetBstSublistMode{f}
\mciteSetBstMaxWidthForm{subitem}{(\alph{mcitesubitemcount})}
\mciteSetBstSublistLabelBeginEnd
  {\mcitemaxwidthsubitemform\space}
  {\relax}
  {\relax}

\bibitem[Armand and Tarascon(2008)Armand, and Tarascon]{Armand2008}
Armand,~M.; Tarascon,~J.-M. Building better batteries. \emph{Nature}
  \textbf{2008}, \emph{451}, 652--657\relax
\mciteBstWouldAddEndPuncttrue
\mciteSetBstMidEndSepPunct{\mcitedefaultmidpunct}
{\mcitedefaultendpunct}{\mcitedefaultseppunct}\relax
\EndOfBibitem
\bibitem[Dunn \latin{et~al.}(2011)Dunn, Kamath, and Tarascon]{Dunn2011}
Dunn,~B.; Kamath,~H.; Tarascon,~J.-M. Electrical Energy Storage for the Grid: A
  Battery of Choices. \emph{Science} \textbf{2011}, \emph{334}, 928--935\relax
\mciteBstWouldAddEndPuncttrue
\mciteSetBstMidEndSepPunct{\mcitedefaultmidpunct}
{\mcitedefaultendpunct}{\mcitedefaultseppunct}\relax
\EndOfBibitem
\bibitem[Islam and Fisher(2014)Islam, and Fisher]{Islam2014}
Islam,~M.~S.; Fisher,~C. A.~J. Lithium and sodium battery cathode materials:
  computational insights into voltage, diffusion and nanostructural properties.
  \emph{Chem. Soc. Rev.} \textbf{2014}, \emph{43}, 185--204\relax
\mciteBstWouldAddEndPuncttrue
\mciteSetBstMidEndSepPunct{\mcitedefaultmidpunct}
{\mcitedefaultendpunct}{\mcitedefaultseppunct}\relax
\EndOfBibitem
\bibitem[Nykvist and Nilsson(2015)Nykvist, and Nilsson]{Nykvist2015}
Nykvist,~B.; Nilsson,~M. Rapidly falling costs of battery packs for electric
  vehicles. \emph{Nat. Clim. Change} \textbf{2015}, \emph{5}, 329--332\relax
\mciteBstWouldAddEndPuncttrue
\mciteSetBstMidEndSepPunct{\mcitedefaultmidpunct}
{\mcitedefaultendpunct}{\mcitedefaultseppunct}\relax
\EndOfBibitem
\bibitem[Janek and Zeier(2016)Janek, and Zeier]{Janek2016}
Janek,~J.; Zeier,~W.~G. A solid future for battery development. \emph{Nat.
  Energy} \textbf{2016}, \emph{1}, 16141\relax
\mciteBstWouldAddEndPuncttrue
\mciteSetBstMidEndSepPunct{\mcitedefaultmidpunct}
{\mcitedefaultendpunct}{\mcitedefaultseppunct}\relax
\EndOfBibitem
\bibitem[Canepa \latin{et~al.}(2017)Canepa, {Sai Gautam}, Hannah, Malik, Liu,
  Gallagher, Persson, and Ceder]{Canepa2017}
Canepa,~P.; {Sai Gautam},~G.; Hannah,~D.~C.; Malik,~R.; Liu,~M.;
  Gallagher,~K.~G.; Persson,~K.~A.; Ceder,~G. Odyssey of Multivalent Cathode
  Materials: Open Questions and Future Challenges. \emph{Chem. Rev.}
  \textbf{2017}, \emph{117}, 4287--4341\relax
\mciteBstWouldAddEndPuncttrue
\mciteSetBstMidEndSepPunct{\mcitedefaultmidpunct}
{\mcitedefaultendpunct}{\mcitedefaultseppunct}\relax
\EndOfBibitem
\bibitem[Thangadurai \latin{et~al.}(2003)Thangadurai, Kaack, and
  Weppner]{Nb2003}
Thangadurai,~V.; Kaack,~H.; Weppner,~W. J.~F. Novel Fast Lithium Ion Conduction
  in Garnet-Type Li$_5$La$_3$M$_2$O$_{12}$ (M$=$Nb, Ta). \emph{J. Am. Ceram.
  Soc.} \textbf{2003}, \emph{40}, 437--440\relax
\mciteBstWouldAddEndPuncttrue
\mciteSetBstMidEndSepPunct{\mcitedefaultmidpunct}
{\mcitedefaultendpunct}{\mcitedefaultseppunct}\relax
\EndOfBibitem
\bibitem[Murugan \latin{et~al.}(2007)Murugan, Thangadurai, and
  Weppner]{Murugan2007}
Murugan,~R.; Thangadurai,~V.; Weppner,~W. Fast lithium ion conduction in
  garnet-type Li$_7$La$_3$Zr$_2$O$_{12}$. \emph{Angew. Chem. Int. Ed.}
  \textbf{2007}, \emph{46}, 7778--7781\relax
\mciteBstWouldAddEndPuncttrue
\mciteSetBstMidEndSepPunct{\mcitedefaultmidpunct}
{\mcitedefaultendpunct}{\mcitedefaultseppunct}\relax
\EndOfBibitem
\bibitem[Knauth(2009)]{Knauth2009}
Knauth,~P. Inorganic solid Li ion conductors: An overview. \emph{Solid State
  Ion.} \textbf{2009}, \emph{180}, 911--916\relax
\mciteBstWouldAddEndPuncttrue
\mciteSetBstMidEndSepPunct{\mcitedefaultmidpunct}
{\mcitedefaultendpunct}{\mcitedefaultseppunct}\relax
\EndOfBibitem
\bibitem[Shinawi and Janek(2013)Shinawi, and Janek]{Shinawi2013}
Shinawi,~H.~E.; Janek,~J. Stabilization of cubic lithium-stuffed garnets of the
  type "Li$_7$La$_3$Zr$_2$O$_{12}$" by addition of gallium. \emph{J. Power
  Sources} \textbf{2013}, \emph{225}, 13--19\relax
\mciteBstWouldAddEndPuncttrue
\mciteSetBstMidEndSepPunct{\mcitedefaultmidpunct}
{\mcitedefaultendpunct}{\mcitedefaultseppunct}\relax
\EndOfBibitem
\bibitem[Thangadurai \latin{et~al.}(2014)Thangadurai, Narayanan, and
  Pinzaru]{Thangadurai2014}
Thangadurai,~V.; Narayanan,~S.; Pinzaru,~D. Garnet-type solid-state fast Li ion
  conductors for Li batteries: critical review. \emph{Chem. Soc. Rev.}
  \textbf{2014}, \emph{43}, 4714--27\relax
\mciteBstWouldAddEndPuncttrue
\mciteSetBstMidEndSepPunct{\mcitedefaultmidpunct}
{\mcitedefaultendpunct}{\mcitedefaultseppunct}\relax
\EndOfBibitem
\bibitem[Kamaya \latin{et~al.}(2011)Kamaya, Homma, Yamakawa, Hirayama, Kanno,
  Yonemura, Kamiyama, Kato, Hama, Kawamoto, and Mitsui]{Kamaya2011}
Kamaya,~N.; Homma,~K.; Yamakawa,~Y.; Hirayama,~M.; Kanno,~R.; Yonemura,~M.;
  Kamiyama,~T.; Kato,~Y.; Hama,~S.; Kawamoto,~K.; Mitsui,~A. A lithium
  superionic conductor. \emph{Nat. Mater.} \textbf{2011}, \emph{10},
  682--686\relax
\mciteBstWouldAddEndPuncttrue
\mciteSetBstMidEndSepPunct{\mcitedefaultmidpunct}
{\mcitedefaultendpunct}{\mcitedefaultseppunct}\relax
\EndOfBibitem
\bibitem[Masquelier(2011)]{Masquelier2011}
Masquelier,~C. Solid electrolytes: Lithium ions on the fast track. \emph{Nat.
  Mater.} \textbf{2011}, \emph{10}, 649--650\relax
\mciteBstWouldAddEndPuncttrue
\mciteSetBstMidEndSepPunct{\mcitedefaultmidpunct}
{\mcitedefaultendpunct}{\mcitedefaultseppunct}\relax
\EndOfBibitem
\bibitem[Wang \latin{et~al.}(2015)Wang, Richards, Ong, Miara, Kim, Mo, and
  Ceder]{Wang2015}
Wang,~Y.; Richards,~W.~D.; Ong,~S.~P.; Miara,~L.~J.; Kim,~J.~C.; Mo,~Y.;
  Ceder,~G. Design principles for solid-state lithium superionic conductors.
  \emph{Nat. Mater.} \textbf{2015}, \emph{14}, 1026--1031\relax
\mciteBstWouldAddEndPuncttrue
\mciteSetBstMidEndSepPunct{\mcitedefaultmidpunct}
{\mcitedefaultendpunct}{\mcitedefaultseppunct}\relax
\EndOfBibitem
\bibitem[Bachman \latin{et~al.}(2016)Bachman, Muy, Grimaud, Chang, Pour, Lux,
  Paschos, Maglia, Lupart, Lamp, Giordano, and Shao-Horn]{Bachman2016}
Bachman,~J.~C.; Muy,~S.; Grimaud,~A.; Chang,~H.-h.; Pour,~N.; Lux,~S.~F.;
  Paschos,~O.; Maglia,~F.; Lupart,~S.; Lamp,~P.; Giordano,~L.; Shao-Horn,~Y.
  Inorganic Solid-State Electrolytes for Lithium Batteries: Mechanisms and
  Properties Governing Ion Conduction. \emph{Chem. Rev.} \textbf{2016},
  \emph{116}, 140--162\relax
\mciteBstWouldAddEndPuncttrue
\mciteSetBstMidEndSepPunct{\mcitedefaultmidpunct}
{\mcitedefaultendpunct}{\mcitedefaultseppunct}\relax
\EndOfBibitem
\bibitem[Kato \latin{et~al.}(2016)Kato, Hori, Saito, Suzuki, Hirayama, Mitsui,
  Yonemura, Iba, and Kanno]{Kato2016}
Kato,~Y.; Hori,~S.; Saito,~T.; Suzuki,~K.; Hirayama,~M.; Mitsui,~A.;
  Yonemura,~M.; Iba,~H.; Kanno,~R. High-power all-solid-state batteries using
  sulfide superionic conductors. \emph{Nat. Energy} \textbf{2016}, \emph{1},
  16030\relax
\mciteBstWouldAddEndPuncttrue
\mciteSetBstMidEndSepPunct{\mcitedefaultmidpunct}
{\mcitedefaultendpunct}{\mcitedefaultseppunct}\relax
\EndOfBibitem
\bibitem[Deng \latin{et~al.}(2015)Deng, Eames, Chotard, Lal{\`{e}}re, Seznec,
  Emge, Pecher, Grey, Masquelier, and Islam]{Deng2015a}
Deng,~Y.; Eames,~C.; Chotard,~J.-N.; Lal{\`{e}}re,~F.; Seznec,~V.; Emge,~S.;
  Pecher,~O.; Grey,~C.~P.; Masquelier,~C.; Islam,~M.~S. Structural and
  mechanistic insights into fast lithium-ion conduction in
  Li$_4$SiO$_4$--Li$_3$PO$_4$ solid electrolytes. \emph{J. Am. Chem. Soc.}
  \textbf{2015}, \emph{137}, 9136--9145\relax
\mciteBstWouldAddEndPuncttrue
\mciteSetBstMidEndSepPunct{\mcitedefaultmidpunct}
{\mcitedefaultendpunct}{\mcitedefaultseppunct}\relax
\EndOfBibitem
\bibitem[Mukhopadhyay \latin{et~al.}(2015)Mukhopadhyay, Thompson, Sakamoto,
  Huq, Wolfenstine, Allen, Bernstein, Stewart, and Johannes]{Mukhopadhyay2015}
Mukhopadhyay,~S.; Thompson,~T.; Sakamoto,~J.; Huq,~A.; Wolfenstine,~J.;
  Allen,~J.~L.; Bernstein,~N.; Stewart,~D.~A.; Johannes,~M.~D. Structure and
  Stoichiometry in Supervalent Doped Li$_7$La$_3$Zr$_2$O$_{12}$. \emph{Chem.
  Mater.} \textbf{2015}, \emph{27}, 3658--3665\relax
\mciteBstWouldAddEndPuncttrue
\mciteSetBstMidEndSepPunct{\mcitedefaultmidpunct}
{\mcitedefaultendpunct}{\mcitedefaultseppunct}\relax
\EndOfBibitem
\bibitem[Deng \latin{et~al.}(2017)Deng, Eames, Fleutot, David, Chotard, Suard,
  Masquelier, and Islam]{Deng2017}
Deng,~Y.; Eames,~C.; Fleutot,~B.; David,~R.; Chotard,~J.-N.; Suard,~E.;
  Masquelier,~C.; Islam,~M.~S. Enhancing the Lithium Ion Conductivity in
  Lithium Superionic Conductor (LISICON) Solid Electrolytes through a Mixed
  Polyanion Effect. \emph{ACS Appl. Mater. Interfaces} \textbf{2017}, \emph{9},
  7050--7058\relax
\mciteBstWouldAddEndPuncttrue
\mciteSetBstMidEndSepPunct{\mcitedefaultmidpunct}
{\mcitedefaultendpunct}{\mcitedefaultseppunct}\relax
\EndOfBibitem
\bibitem[Lotsch and Maier(2017)Lotsch, and Maier]{Lotsch2017}
Lotsch,~B.~V.; Maier,~J. Relevance of solid electrolytes for lithium-based
  batteries: A realistic view. \emph{J. Electroceram.} \textbf{2017},
  \emph{38}, 128--141\relax
\mciteBstWouldAddEndPuncttrue
\mciteSetBstMidEndSepPunct{\mcitedefaultmidpunct}
{\mcitedefaultendpunct}{\mcitedefaultseppunct}\relax
\EndOfBibitem
\bibitem[Kim \latin{et~al.}(2017)Kim, Yoon, Park, and Kang]{Kim2017}
Kim,~J.-J.; Yoon,~K.; Park,~I.; Kang,~K. Progress in the Development of
  Sodium-Ion Solid Electrolytes. \emph{Small Methods} \textbf{2017}, \emph{1},
  1700219\relax
\mciteBstWouldAddEndPuncttrue
\mciteSetBstMidEndSepPunct{\mcitedefaultmidpunct}
{\mcitedefaultendpunct}{\mcitedefaultseppunct}\relax
\EndOfBibitem
\bibitem[Cheng \latin{et~al.}(2013)Cheng, Xu, Ding, Sang, Liu, and
  Cao]{Cheng2013}
Cheng,~G.; Xu,~Q.; Ding,~F.; Sang,~L.; Liu,~X.; Cao,~D. {Electrochemical
  behavior of aluminum in Grignard reagents/THF electrolytic solutions for
  rechargeable magnesium batteries}. \emph{Electrochimica Acta} \textbf{2013},
  \emph{88}, 790--797\relax
\mciteBstWouldAddEndPuncttrue
\mciteSetBstMidEndSepPunct{\mcitedefaultmidpunct}
{\mcitedefaultendpunct}{\mcitedefaultseppunct}\relax
\EndOfBibitem
\bibitem[Buschmann \latin{et~al.}(2011)Buschmann, D{\"{o}}lle, Berendts, Kuhn,
  Bottke, Wilkening, Heitjans, Senyshyn, Ehrenberg, Lotnyk, Duppel, Kienle, and
  Janek]{Buschmann2011}
Buschmann,~H.; D{\"{o}}lle,~J.; Berendts,~S.; Kuhn,~A.; Bottke,~P.;
  Wilkening,~M.; Heitjans,~P.; Senyshyn,~A.; Ehrenberg,~H.; Lotnyk,~A.;
  Duppel,~V.; Kienle,~L.; Janek,~J. Structure and dynamics of the fast lithium
  ion conductor “Li$_7$La$_3$Zr$_2$O$_{12}$”. \emph{Phys. Chem. Chem.
  Phys.} \textbf{2011}, \emph{13}, 19378--19392\relax
\mciteBstWouldAddEndPuncttrue
\mciteSetBstMidEndSepPunct{\mcitedefaultmidpunct}
{\mcitedefaultendpunct}{\mcitedefaultseppunct}\relax
\EndOfBibitem
\bibitem[Luntz \latin{et~al.}(2015)Luntz, Voss, and Reuter]{Luntz2015}
Luntz,~A.~C.; Voss,~J.; Reuter,~K. Interfacial Challenges in Solid-State Li Ion
  Batteries. \emph{J. Phys. Chem. Lett.} \textbf{2015}, \emph{6},
  4599--4604\relax
\mciteBstWouldAddEndPuncttrue
\mciteSetBstMidEndSepPunct{\mcitedefaultmidpunct}
{\mcitedefaultendpunct}{\mcitedefaultseppunct}\relax
\EndOfBibitem
\bibitem[Richards \latin{et~al.}(2016)Richards, Miara, Wang, Kim, and
  Ceder]{Richards2016}
Richards,~W.~D.; Miara,~L.~J.; Wang,~Y.; Kim,~J.~C.; Ceder,~G. Interface
  Stability in Solid-State Batteries. \emph{Chem. Mater.} \textbf{2016},
  \emph{28}, 266--273\relax
\mciteBstWouldAddEndPuncttrue
\mciteSetBstMidEndSepPunct{\mcitedefaultmidpunct}
{\mcitedefaultendpunct}{\mcitedefaultseppunct}\relax
\EndOfBibitem
\bibitem[Yu \latin{et~al.}(2016)Yu, Ganapathy, de~Klerk, Roslon, van Eck,
  Kentgens, and Wagemaker]{Yu2016b}
Yu,~C.; Ganapathy,~S.; de~Klerk,~N. J.~J.; Roslon,~I.; van Eck,~E. R.~H.;
  Kentgens,~A. P.~M.; Wagemaker,~M. Unravelling Li-Ion Transport from
  Picoseconds to Seconds: Bulk versus Interfaces in an Argyrodite
  Li$_6$PS$_5$Cl–Li$_2$S All-Solid-State Li-Ion Battery. \emph{J. Am. Chem.
  Soc.} \textbf{2016}, \emph{138}, 11192--11201\relax
\mciteBstWouldAddEndPuncttrue
\mciteSetBstMidEndSepPunct{\mcitedefaultmidpunct}
{\mcitedefaultendpunct}{\mcitedefaultseppunct}\relax
\EndOfBibitem
\bibitem[Kerman \latin{et~al.}(2017)Kerman, Luntz, Viswanathan, Chiang, and
  Chen]{Kerman2017}
Kerman,~K.; Luntz,~A.; Viswanathan,~V.; Chiang,~Y.-M.; Chen,~Z.
  Review—Practical Challenges Hindering the Development of Solid State Li Ion
  Batteries. \emph{J. Electrochem. Soc.} \textbf{2017}, \emph{164},
  A1731--A1744\relax
\mciteBstWouldAddEndPuncttrue
\mciteSetBstMidEndSepPunct{\mcitedefaultmidpunct}
{\mcitedefaultendpunct}{\mcitedefaultseppunct}\relax
\EndOfBibitem
\bibitem[Yonemoto \latin{et~al.}(2017)Yonemoto, Nishimura, Motoyama,
  Tsuchimine, Kobayashi, and Iriyama]{Yonemoto2017}
Yonemoto,~F.; Nishimura,~A.; Motoyama,~M.; Tsuchimine,~N.; Kobayashi,~S.;
  Iriyama,~Y. Temperature effects on cycling stability of Li plating/stripping
  on Ta-doped Li$_7$La$_3$Zr$_2$O$_{12}$. \emph{J. Power Sources}
  \textbf{2017}, \emph{343}, 207--215\relax
\mciteBstWouldAddEndPuncttrue
\mciteSetBstMidEndSepPunct{\mcitedefaultmidpunct}
{\mcitedefaultendpunct}{\mcitedefaultseppunct}\relax
\EndOfBibitem
\bibitem[Porz \latin{et~al.}(2017)Porz, Swamy, Sheldon, Rettenwander,
  Fr{\"{o}}mling, Thaman, Berendts, Uecker, Carter, and Chiang]{Porz2017}
Porz,~L.; Swamy,~T.; Sheldon,~B.~W.; Rettenwander,~D.; Fr{\"{o}}mling,~T.;
  Thaman,~H.~L.; Berendts,~S.; Uecker,~R.; Carter,~W.~C.; Chiang,~Y.-M.
  Mechanism of Lithium Metal Penetration through Inorganic Solid Electrolytes.
  \emph{Adv. Energy Mater.} \textbf{2017}, 1701003\relax
\mciteBstWouldAddEndPuncttrue
\mciteSetBstMidEndSepPunct{\mcitedefaultmidpunct}
{\mcitedefaultendpunct}{\mcitedefaultseppunct}\relax
\EndOfBibitem
\bibitem[Hanft \latin{et~al.}(2017)Hanft, Exner, and Moos]{Hanft2017}
Hanft,~D.; Exner,~J.; Moos,~R. Thick-films of garnet-type lithium ion conductor
  prepared by the Aerosol Deposition Method: The role of morphology and
  annealing treatment on the ionic conductivity. \emph{J. Power Sources}
  \textbf{2017}, \emph{361}, 61--69\relax
\mciteBstWouldAddEndPuncttrue
\mciteSetBstMidEndSepPunct{\mcitedefaultmidpunct}
{\mcitedefaultendpunct}{\mcitedefaultseppunct}\relax
\EndOfBibitem
\bibitem[Geiger \latin{et~al.}(2011)Geiger, Alekseev, Lazic, Fisch, Armbruster,
  Langner, Fechtelkord, Kim, Pettke, and Weppner]{Geiger2011}
Geiger,~C.~A.; Alekseev,~E.; Lazic,~B.; Fisch,~M.; Armbruster,~T.; Langner,~R.;
  Fechtelkord,~M.; Kim,~N.; Pettke,~T.; Weppner,~W. Crystal Chemistry and
  Stability of Li$_7$La$_3$Zr$_2$O$_{12}$ Garnet: A Fast Lithium-Ion Conductor.
  \emph{Inorg. Chem.} \textbf{2011}, \emph{50}, 1089--1097\relax
\mciteBstWouldAddEndPuncttrue
\mciteSetBstMidEndSepPunct{\mcitedefaultmidpunct}
{\mcitedefaultendpunct}{\mcitedefaultseppunct}\relax
\EndOfBibitem
\bibitem[Kuhn \latin{et~al.}(2011)Kuhn, Narayanan, Spencer, Goward,
  Thangadurai, and Wilkening]{Kuhn2011}
Kuhn,~A.; Narayanan,~S.; Spencer,~L.; Goward,~G.; Thangadurai,~V.;
  Wilkening,~M. Li self-diffusion in garnet-type Li$_7$La$_3$Zr$_2$O$_{12}$ as
  probed directly by diffusion-induced $^{7}$Li spin-lattice relaxation NMR
  spectroscopy. \emph{Phys. Rev. B} \textbf{2011}, \emph{83}, 094302\relax
\mciteBstWouldAddEndPuncttrue
\mciteSetBstMidEndSepPunct{\mcitedefaultmidpunct}
{\mcitedefaultendpunct}{\mcitedefaultseppunct}\relax
\EndOfBibitem
\bibitem[Allen \latin{et~al.}(2012)Allen, Wolfenstine, Rangasamy, and
  Sakamoto]{Allen2012}
Allen,~J.~L.; Wolfenstine,~J.; Rangasamy,~E.; Sakamoto,~J. Effect of
  substitution (Ta, Al, Ga) on the conductivity of Li$_7$La$_3$Zr$_2$O$_{12}$.
  \emph{J. Power Sources} \textbf{2012}, \emph{206}, 315--319\relax
\mciteBstWouldAddEndPuncttrue
\mciteSetBstMidEndSepPunct{\mcitedefaultmidpunct}
{\mcitedefaultendpunct}{\mcitedefaultseppunct}\relax
\EndOfBibitem
\bibitem[Adams and Rao(2012)Adams, and Rao]{Adams2012a}
Adams,~S.; Rao,~R.~P. Ion transport and phase transition in
  Li$_{7-x}$La$_3$(Zr$_{2-x}$M$_x$)O$_{12}$ (M = Ta$^{5+}$, Nb$^{5+}$, x = 0,
  0.25). \emph{J. Mater. Chem.} \textbf{2012}, \emph{22}, 1426--1434\relax
\mciteBstWouldAddEndPuncttrue
\mciteSetBstMidEndSepPunct{\mcitedefaultmidpunct}
{\mcitedefaultendpunct}{\mcitedefaultseppunct}\relax
\EndOfBibitem
\bibitem[Sharafi \latin{et~al.}(2017)Sharafi, Haslam, Kerns, Wolfenstine, and
  Sakamoto]{Sharafi2017}
Sharafi,~A.; Haslam,~C.~G.; Kerns,~R.~D.; Wolfenstine,~J.; Sakamoto,~J.
  Controlling and correlating the effect of grain size with the mechanical and
  electrochemical properties of Li7La3Zr2O12 solid-state electrolyte. \emph{J.
  Mater. Chem. A} \textbf{2017}, \emph{5}, 21491--21504\relax
\mciteBstWouldAddEndPuncttrue
\mciteSetBstMidEndSepPunct{\mcitedefaultmidpunct}
{\mcitedefaultendpunct}{\mcitedefaultseppunct}\relax
\EndOfBibitem
\bibitem[Yi \latin{et~al.}(2017)Yi, Wang, Kieffer, and Laine]{Yi2017}
Yi,~E.; Wang,~W.; Kieffer,~J.; Laine,~R.~M. Key parameters governing the
  densification of cubic-Li$_7$La$_3$Zr$_2$O$_{12}$ Li$^{+}$ conductors.
  \emph{J. Power Sources} \textbf{2017}, \emph{352}, 156--164\relax
\mciteBstWouldAddEndPuncttrue
\mciteSetBstMidEndSepPunct{\mcitedefaultmidpunct}
{\mcitedefaultendpunct}{\mcitedefaultseppunct}\relax
\EndOfBibitem
\bibitem[Cheng \latin{et~al.}(2015)Cheng, Wu, Jarry, Chen, Ye, Zhu, Kostecki,
  Persson, Guo, Salmeron, Chen, and Doeff]{Cheng2015a}
Cheng,~L.; Wu,~C.~H.; Jarry,~A.; Chen,~W.; Ye,~Y.; Zhu,~J.; Kostecki,~R.;
  Persson,~K.; Guo,~J.; Salmeron,~M.; Chen,~G.; Doeff,~M. Interrelationships
  among Grain Size, Surface Composition, Air Stability, and Interfacial
  Resistance of Al-Substituted Li$_7$La$_3$Zr$_2$O$_{12}$ Solid Electrolytes.
  \emph{ACS Appl. Mater. Interfaces} \textbf{2015}, \emph{7},
  17649--17655\relax
\mciteBstWouldAddEndPuncttrue
\mciteSetBstMidEndSepPunct{\mcitedefaultmidpunct}
{\mcitedefaultendpunct}{\mcitedefaultseppunct}\relax
\EndOfBibitem
\bibitem[Ma \latin{et~al.}(2016)Ma, Cheng, Yin, Luo, Sharafi, Sakamoto, Li,
  More, Dudney, and Chi]{Ma2016}
Ma,~C.; Cheng,~Y.; Yin,~K.; Luo,~J.; Sharafi,~A.; Sakamoto,~J.; Li,~J.;
  More,~K.~L.; Dudney,~N.~J.; Chi,~M. Interfacial Stability of Li Metal–Solid
  Electrolyte Elucidated via in Situ Electron Microscopy. \emph{Nano Lett.}
  \textbf{2016}, \emph{16}, 7030--7036\relax
\mciteBstWouldAddEndPuncttrue
\mciteSetBstMidEndSepPunct{\mcitedefaultmidpunct}
{\mcitedefaultendpunct}{\mcitedefaultseppunct}\relax
\EndOfBibitem
\bibitem[Han \latin{et~al.}(2016)Han, Gong, Fu, He, Hitz, Dai, Pearse, Liu,
  Wang, Rubloff, Mo, Thangadurai, Wachsman, and Hu]{Han2016}
Han,~X.; Gong,~Y.; Fu,~K.~K.; He,~X.; Hitz,~G.~T.; Dai,~J.; Pearse,~A.;
  Liu,~B.; Wang,~H.; Rubloff,~G.; Mo,~Y.; Thangadurai,~V.; Wachsman,~E.~D.;
  Hu,~L. Negating interfacial impedance in garnet-based solid-state Li metal
  batteries. \emph{Nat. Mater.} \textbf{2016}, \emph{16}, 572--579\relax
\mciteBstWouldAddEndPuncttrue
\mciteSetBstMidEndSepPunct{\mcitedefaultmidpunct}
{\mcitedefaultendpunct}{\mcitedefaultseppunct}\relax
\EndOfBibitem
\bibitem[Kim \latin{et~al.}(2016)Kim, Yoo, Schmidt, Sharafi, Lee, Wolfenstine,
  and Sakamoto]{Kim2016}
Kim,~Y.; Yoo,~A.; Schmidt,~R.; Sharafi,~A.; Lee,~H.; Wolfenstine,~J.;
  Sakamoto,~J. Electrochemical Stability of
  Li$_{6.5}$La$_3$Zr$_{1.5}$M$_{0.5}$O$_{12}$ (M = Nb or Ta) against Metallic
  Lithium. \emph{Front. Energy Res.} \textbf{2016}, \emph{4}, 1--7\relax
\mciteBstWouldAddEndPuncttrue
\mciteSetBstMidEndSepPunct{\mcitedefaultmidpunct}
{\mcitedefaultendpunct}{\mcitedefaultseppunct}\relax
\EndOfBibitem
\bibitem[Wang \latin{et~al.}(2017)Wang, Gong, Dai, Zhang, Xie, Pastel, Liu,
  Wachsman, Wang, and Hu]{Wang2017}
Wang,~C.; Gong,~Y.; Dai,~J.; Zhang,~L.; Xie,~H.; Pastel,~G.; Liu,~B.;
  Wachsman,~E.; Wang,~H.; Hu,~L. In Situ Neutron Depth Profiling of Lithium
  Metal–Garnet Interfaces for Solid State Batteries. \emph{J. Am. Chem. Soc.}
  \textbf{2017}, \emph{139}, 14257--14264\relax
\mciteBstWouldAddEndPuncttrue
\mciteSetBstMidEndSepPunct{\mcitedefaultmidpunct}
{\mcitedefaultendpunct}{\mcitedefaultseppunct}\relax
\EndOfBibitem
\bibitem[Kotobuki \latin{et~al.}(2010)Kotobuki, Munakata, Kanamura, Sato, and
  Yoshida]{Kotobuki2010}
Kotobuki,~M.; Munakata,~H.; Kanamura,~K.; Sato,~Y.; Yoshida,~T. Compatibility
  of Li$_7$La$_3$Zr$_2$O$_{12}$ Solid Electrolyte to All-Solid-State Battery
  Using Li Metal Anode. \emph{J. Electrochem. Soc.} \textbf{2010}, \emph{157},
  A1076--A1079\relax
\mciteBstWouldAddEndPuncttrue
\mciteSetBstMidEndSepPunct{\mcitedefaultmidpunct}
{\mcitedefaultendpunct}{\mcitedefaultseppunct}\relax
\EndOfBibitem
\bibitem[Sharafi \latin{et~al.}(2016)Sharafi, Meyer, Nanda, Wolfenstine, and
  Sakamoto]{Sharafi2016}
Sharafi,~A.; Meyer,~H.~M.; Nanda,~J.; Wolfenstine,~J.; Sakamoto,~J.
  Characterizing the Li-Li$_7$La$_3$Zr$_2$O$_{12}$ interface stability and
  kinetics as a function of temperature and current density. \emph{J. Power
  Sources} \textbf{2016}, \emph{302}, 135--139\relax
\mciteBstWouldAddEndPuncttrue
\mciteSetBstMidEndSepPunct{\mcitedefaultmidpunct}
{\mcitedefaultendpunct}{\mcitedefaultseppunct}\relax
\EndOfBibitem
\bibitem[Wang \latin{et~al.}(2017)Wang, Gong, Dai, Zhang, Xie, Pastel, Liu,
  Wachsman, Wang, and Hu]{Wang2017a}
Wang,~C.; Gong,~Y.; Dai,~J.; Zhang,~L.; Xie,~H.; Pastel,~G.; Liu,~B.;
  Wachsman,~E.; Wang,~H.; Hu,~L. In Situ Neutron Depth Profiling of Lithium
  Metal-Garnet Interfaces for Solid State Batteries. \emph{J. Am. Chem. Soc.}
  \textbf{2017}, \emph{139}, 14257--14264\relax
\mciteBstWouldAddEndPuncttrue
\mciteSetBstMidEndSepPunct{\mcitedefaultmidpunct}
{\mcitedefaultendpunct}{\mcitedefaultseppunct}\relax
\EndOfBibitem
\bibitem[Kingon and Clark(1983)Kingon, and Clark]{Kingon1983}
Kingon,~A.~I.; Clark,~J.~B. Sintering of PZT Ceramics: I, Atmosphere Control.
  \emph{J. Am. Ceram. Soc.} \textbf{1983}, \emph{66}, 253--256\relax
\mciteBstWouldAddEndPuncttrue
\mciteSetBstMidEndSepPunct{\mcitedefaultmidpunct}
{\mcitedefaultendpunct}{\mcitedefaultseppunct}\relax
\EndOfBibitem
\bibitem[Awaka \latin{et~al.}(2009)Awaka, Kijima, Hayakawa, and
  Akimoto]{Awaka2009}
Awaka,~J.; Kijima,~N.; Hayakawa,~H.; Akimoto,~J. Synthesis and structure
  analysis of tetragonal Li$_7$La$_3$Zr$_2$O$_{12}$ with the garnet-related
  type structure. \emph{J. Solid State Chem.} \textbf{2009}, \emph{182},
  2046--2052\relax
\mciteBstWouldAddEndPuncttrue
\mciteSetBstMidEndSepPunct{\mcitedefaultmidpunct}
{\mcitedefaultendpunct}{\mcitedefaultseppunct}\relax
\EndOfBibitem
\bibitem[Miara \latin{et~al.}(2013)Miara, Ong, Mo, Richards, Park, Lee, Lee,
  and Ceder]{Miara2013}
Miara,~L.~J.; Ong,~S.~P.; Mo,~Y.; Richards,~W.~D.; Park,~Y.; Lee,~J.-M.;
  Lee,~H.~S.; Ceder,~G. Effect of Rb and Ta Doping on the Ionic Conductivity
  and Stability of the Garnet Li$_{7+2x – y}$(La$_{3– x}$Rb$_x$)(Zr$_{2–
  y}$Ta$_y$)O$_{12}$ (0 $\leq$ x $\leq$ 0.375, 0 $\leq$ y $\leq$ 1) Superionic
  Conductor: A First Principles Investigation. \emph{Chem. Mater.}
  \textbf{2013}, \emph{25}, 3048--3055\relax
\mciteBstWouldAddEndPuncttrue
\mciteSetBstMidEndSepPunct{\mcitedefaultmidpunct}
{\mcitedefaultendpunct}{\mcitedefaultseppunct}\relax
\EndOfBibitem
\bibitem[Thompson \latin{et~al.}(2017)Thompson, Yu, Williams, Schmidt,
  Garcia-Mendez, Wolfenstine, Allen, Kioupakis, Siegel, and
  Sakamoto]{Thompson2017}
Thompson,~T.; Yu,~S.; Williams,~L.; Schmidt,~R.~D.; Garcia-Mendez,~R.;
  Wolfenstine,~J.; Allen,~J.~L.; Kioupakis,~E.; Siegel,~D.~J.; Sakamoto,~J.
  Electrochemical Window of the Li-Ion Solid Electrolyte
  Li$_7$La$_3$Zr$_2$O$_{12}$. \emph{ACS Energy Lett.} \textbf{2017}, \emph{2},
  462--468\relax
\mciteBstWouldAddEndPuncttrue
\mciteSetBstMidEndSepPunct{\mcitedefaultmidpunct}
{\mcitedefaultendpunct}{\mcitedefaultseppunct}\relax
\EndOfBibitem
\bibitem[Jain \latin{et~al.}(2013)Jain, Ong, Hautier, Chen, Richards, Dacek,
  Cholia, Gunter, Skinner, Ceder, and Persson]{Jain2013}
Jain,~A.; Ong,~S.~P.; Hautier,~G.; Chen,~W.; Richards,~W.~D.; Dacek,~S.;
  Cholia,~S.; Gunter,~D.; Skinner,~D.; Ceder,~G.; Persson,~K.~A. Commentary:
  The Materials Project: A materials genome approach to accelerating materials
  innovation. \emph{APL Mater.} \textbf{2013}, \emph{1}, 011002\relax
\mciteBstWouldAddEndPuncttrue
\mciteSetBstMidEndSepPunct{\mcitedefaultmidpunct}
{\mcitedefaultendpunct}{\mcitedefaultseppunct}\relax
\EndOfBibitem
\bibitem[Chen \latin{et~al.}(2015)Chen, Puchala, and der Ven]{Chen2015a}
Chen,~M.-H.; Puchala,~B.; der Ven,~A.~V. High-temperature stability of
  $\delta'$-ZrO. \emph{Calphad} \textbf{2015}, \emph{51}, 292--298\relax
\mciteBstWouldAddEndPuncttrue
\mciteSetBstMidEndSepPunct{\mcitedefaultmidpunct}
{\mcitedefaultendpunct}{\mcitedefaultseppunct}\relax
\EndOfBibitem
\bibitem[Burbano \latin{et~al.}(2016)Burbano, Carlier, Boucher, Morgan, and
  Salanne]{Burbano2016}
Burbano,~M.; Carlier,~D.; Boucher,~F.; Morgan,~B.~J.; Salanne,~M. Sparse Cyclic
  Excitations Explain the Low Ionic Conductivity of
  StoichiometricLi$_7$La$_3$Zr$_2$O$_{12}$. \emph{Phys. Rev. Lett.}
  \textbf{2016}, \emph{116}\relax
\mciteBstWouldAddEndPuncttrue
\mciteSetBstMidEndSepPunct{\mcitedefaultmidpunct}
{\mcitedefaultendpunct}{\mcitedefaultseppunct}\relax
\EndOfBibitem
\bibitem[Haase and Sauer(1998)Haase, and Sauer]{Haase1998}
Haase,~F.; Sauer,~J. The Surface Structure of Sulfated Zirconia:~ Periodic ab
  Initio Study of Sulfuric Acid Adsorbed on ZrO$_2$(101) and ZrO$_2$(001).
  \emph{J. Am. Chem. Soc.} \textbf{1998}, \emph{120}, 13503--13512\relax
\mciteBstWouldAddEndPuncttrue
\mciteSetBstMidEndSepPunct{\mcitedefaultmidpunct}
{\mcitedefaultendpunct}{\mcitedefaultseppunct}\relax
\EndOfBibitem
\bibitem[Reuter and Scheffler(2001)Reuter, and Scheffler]{Reuter2001}
Reuter,~K.; Scheffler,~M. Composition, structure, and stability of RuO$_2$
  (110) as a function of oxygen pressure. \emph{Phys. Rev. B} \textbf{2001},
  \emph{65}, 035406\relax
\mciteBstWouldAddEndPuncttrue
\mciteSetBstMidEndSepPunct{\mcitedefaultmidpunct}
{\mcitedefaultendpunct}{\mcitedefaultseppunct}\relax
\EndOfBibitem
\bibitem[Antoini(1992)]{Antoini1992}
Antoini,~E. Sintering of Li$_x$Ni$_{1-x}$O solid solutions at 1200 $^{\circ}$C.
  \emph{J. Mater. Sci.} \textbf{1992}, \emph{27}, 3335--3340\relax
\mciteBstWouldAddEndPuncttrue
\mciteSetBstMidEndSepPunct{\mcitedefaultmidpunct}
{\mcitedefaultendpunct}{\mcitedefaultseppunct}\relax
\EndOfBibitem
\bibitem[Li \latin{et~al.}(2013)Li, Cao, and Guo]{Li_2013}
Li,~Y.; Cao,~Y.; Guo,~X. Influence of lithium oxide additives on densification
  and ionic conductivity of garnet-type
  Li$_{6.75}$La$_3$Zr$_{1.75}$Ta$_{0.25}$O$_{12}$ solid electrolytes.
  \emph{Solid State Ionics} \textbf{2013}, \emph{253}, 76--80\relax
\mciteBstWouldAddEndPuncttrue
\mciteSetBstMidEndSepPunct{\mcitedefaultmidpunct}
{\mcitedefaultendpunct}{\mcitedefaultseppunct}\relax
\EndOfBibitem
\bibitem[Brown(1988)]{Brown1988}
Brown,~I.~D. What factors determine cation coordination numbers? \emph{Acta
  Crystallogr. B} \textbf{1988}, \emph{44}, 545--553\relax
\mciteBstWouldAddEndPuncttrue
\mciteSetBstMidEndSepPunct{\mcitedefaultmidpunct}
{\mcitedefaultendpunct}{\mcitedefaultseppunct}\relax
\EndOfBibitem
\bibitem[Rong \latin{et~al.}(2015)Rong, Malik, Canepa, {Sai Gautam}, Liu, Jain,
  Persson, and Ceder]{Rong2015}
Rong,~Z.; Malik,~R.; Canepa,~P.; {Sai Gautam},~G.; Liu,~M.; Jain,~A.;
  Persson,~K.; Ceder,~G. Materials Design Rules for Multivalent Ion Mobility in
  Intercalation Structures. \emph{Chem. Mater.} \textbf{2015}, \emph{27},
  6016--6021\relax
\mciteBstWouldAddEndPuncttrue
\mciteSetBstMidEndSepPunct{\mcitedefaultmidpunct}
{\mcitedefaultendpunct}{\mcitedefaultseppunct}\relax
\EndOfBibitem
\bibitem[Kubicek \latin{et~al.}(2017)Kubicek, Wachter-Welzl, Rettenwander,
  Wagner, Berendts, Uecker, Amthauer, Hutter, and Fleig]{Kubicek2017}
Kubicek,~M.; Wachter-Welzl,~A.; Rettenwander,~D.; Wagner,~R.; Berendts,~S.;
  Uecker,~R.; Amthauer,~G.; Hutter,~H.; Fleig,~J. Oxygen Vacancies in Fast
  Lithium-Ion Conducting Garnets. \emph{Chem. Mater.} \textbf{2017}, \emph{29},
  7189--7196\relax
\mciteBstWouldAddEndPuncttrue
\mciteSetBstMidEndSepPunct{\mcitedefaultmidpunct}
{\mcitedefaultendpunct}{\mcitedefaultseppunct}\relax
\EndOfBibitem
\bibitem[Cheng \latin{et~al.}(2014)Cheng, Park, Hou, Zorba, Chen, Richardson,
  Cabana, Russo, and Doeff]{Cheng2014}
Cheng,~L.; Park,~J.~S.; Hou,~H.; Zorba,~V.; Chen,~G.; Richardson,~T.;
  Cabana,~J.; Russo,~R.; Doeff,~M. Effect of microstructure and surface
  impurity segregation on the electrical and electrochemical properties of
  dense Al-substituted Li$_7$La$_3$Zr$_2$O$_{12}$. \emph{J. Mater. Chem. A}
  \textbf{2014}, \emph{2}, 172--181\relax
\mciteBstWouldAddEndPuncttrue
\mciteSetBstMidEndSepPunct{\mcitedefaultmidpunct}
{\mcitedefaultendpunct}{\mcitedefaultseppunct}\relax
\EndOfBibitem
\bibitem[Ohta \latin{et~al.}(2016)Ohta, Kihira, and Asaoka]{Ohta2016}
Ohta,~S.; Kihira,~Y.; Asaoka,~T. Grain Boundary Analysis of the Garnet-Like
  Oxides Li$_{7+x-y}$La$_{3-x}$AXZr$_{2-y}$Nb$_Y$O$_{12}$ (A = Sr or Ca).
  \emph{Front. Energy Res.} \textbf{2016}, \emph{4}\relax
\mciteBstWouldAddEndPuncttrue
\mciteSetBstMidEndSepPunct{\mcitedefaultmidpunct}
{\mcitedefaultendpunct}{\mcitedefaultseppunct}\relax
\EndOfBibitem
\bibitem[Jamnik and Maier(1999)Jamnik, and Maier]{Jamnik1999}
Jamnik,~J.; Maier,~J. Treatment of the Impedance of Mixed Conductors Equivalent
  Circuit Model and Explicit Approximate Solutions. \emph{J. Electrochem. Soc.}
  \textbf{1999}, \emph{146}, 4183--4188\relax
\mciteBstWouldAddEndPuncttrue
\mciteSetBstMidEndSepPunct{\mcitedefaultmidpunct}
{\mcitedefaultendpunct}{\mcitedefaultseppunct}\relax
\EndOfBibitem
\bibitem[Jamnik and Maier(2001)Jamnik, and Maier]{Jamnik2001}
Jamnik,~J.; Maier,~J. Generalised equivalent circuits for mass and charge
  transport: chemical capacitance and its implications. \emph{Phys. Chem. Chem.
  Phys.} \textbf{2001}, \emph{3}, 1668--1678\relax
\mciteBstWouldAddEndPuncttrue
\mciteSetBstMidEndSepPunct{\mcitedefaultmidpunct}
{\mcitedefaultendpunct}{\mcitedefaultseppunct}\relax
\EndOfBibitem
\bibitem[Lai and Haile(2005)Lai, and Haile]{Lai2005}
Lai,~W.; Haile,~S.~M. Impedance spectroscopy as a tool for chemical and
  electrochemical analysis of mixed conductors: A case study of ceria. \emph{J.
  Am. Ceram. Soc.} \textbf{2005}, \emph{88}, 2979--2997\relax
\mciteBstWouldAddEndPuncttrue
\mciteSetBstMidEndSepPunct{\mcitedefaultmidpunct}
{\mcitedefaultendpunct}{\mcitedefaultseppunct}\relax
\EndOfBibitem
\bibitem[Dawson \latin{et~al.}(2017)Dawson, Canepa, Famprikis, Masquelier, and
  Islam]{Dawson2017}
Dawson,~J.~A.; Canepa,~P.; Famprikis,~T.; Masquelier,~C.; Islam,~M.~S.
  Atomic-Scale Influence of Grain Boundaries on Li-Ion Conduction in Solid
  Electrolytes for All-Solid-State Batteries. \emph{J. Am. Chem. Soc.}
  \textbf{2017}, \emph{140}, 362--368\relax
\mciteBstWouldAddEndPuncttrue
\mciteSetBstMidEndSepPunct{\mcitedefaultmidpunct}
{\mcitedefaultendpunct}{\mcitedefaultseppunct}\relax
\EndOfBibitem
\bibitem[Kazyak \latin{et~al.}(2017)Kazyak, Chen, Wood, Davis, Thompson,
  Bielinski, Sanchez, Wang, Wang, Sakamoto, and Dasgupta]{Kazyak2017}
Kazyak,~E.; Chen,~K.-H.; Wood,~K.~N.; Davis,~A.~L.; Thompson,~T.;
  Bielinski,~A.~R.; Sanchez,~A.~J.; Wang,~X.; Wang,~C.; Sakamoto,~J.;
  Dasgupta,~N.~P. Atomic Layer Deposition of the Solid Electrolyte Garnet
  Li$_7$La$_3$Zr$_2$O$_{12}$. \emph{Chem. Mater.} \textbf{2017}, \emph{29},
  3785--3792\relax
\mciteBstWouldAddEndPuncttrue
\mciteSetBstMidEndSepPunct{\mcitedefaultmidpunct}
{\mcitedefaultendpunct}{\mcitedefaultseppunct}\relax
\EndOfBibitem
\bibitem[Amores \latin{et~al.}(2016)Amores, Ashton, Baker, Cussen, and
  Corr]{Amores2016}
Amores,~M.; Ashton,~T.~E.; Baker,~P.~J.; Cussen,~E.~J.; Corr,~S.~A. Fast
  microwave-assisted synthesis of Li-stuffed garnets and insights into Li
  diffusion from muon spin spectroscopy. \emph{J. Mater. Chem.A} \textbf{2016},
  \emph{4}, 1729--1736\relax
\mciteBstWouldAddEndPuncttrue
\mciteSetBstMidEndSepPunct{\mcitedefaultmidpunct}
{\mcitedefaultendpunct}{\mcitedefaultseppunct}\relax
\EndOfBibitem
\bibitem[Miara \latin{et~al.}(2015)Miara, Richards, Wang, and Ceder]{Miara2015}
Miara,~L.~J.; Richards,~W.~D.; Wang,~Y.~E.; Ceder,~G. {First-principles studies
  on cation dopants and electrolyte|cathode interphases for lithium garnets}.
  \emph{Chem. Mater.} \textbf{2015}, \emph{27}, 4040--4047\relax
\mciteBstWouldAddEndPuncttrue
\mciteSetBstMidEndSepPunct{\mcitedefaultmidpunct}
{\mcitedefaultendpunct}{\mcitedefaultseppunct}\relax
\EndOfBibitem
\bibitem[Park \latin{et~al.}(2016)Park, Yu, Jung, Li, Zhou, Gao, Son, and
  Goodenough]{Park2016}
Park,~K.; Yu,~B.-C.; Jung,~J.-W.; Li,~Y.; Zhou,~W.; Gao,~H.; Son,~S.;
  Goodenough,~J.~B. Electrochemical Nature of the Cathode Interface for a
  Solid-State Lithium-Ion Battery: Interface between LiCoO$_2$ and
  Garnet-Li$_7$La$_3$Zr$_2$O$_{12}$. \emph{Chem. Mater.} \textbf{2016},
  \emph{28}, 8051--8059\relax
\mciteBstWouldAddEndPuncttrue
\mciteSetBstMidEndSepPunct{\mcitedefaultmidpunct}
{\mcitedefaultendpunct}{\mcitedefaultseppunct}\relax
\EndOfBibitem
\bibitem[Chase(1998)]{Chase1996}
Chase,~M.~W. \emph{NIST-JANAF Thermochemical Tables 2 Volume-Set}; American
  Institute of Physics, 1998; p 1963\relax
\mciteBstWouldAddEndPuncttrue
\mciteSetBstMidEndSepPunct{\mcitedefaultmidpunct}
{\mcitedefaultendpunct}{\mcitedefaultseppunct}\relax
\EndOfBibitem
\bibitem[Tasker(1979)]{Tasker1979}
Tasker,~P.~W. The stability of ionic crystal surfaces. \emph{J. Phys. C: Solid
  State Phys.} \textbf{1979}, \emph{12}, 4977--4984\relax
\mciteBstWouldAddEndPuncttrue
\mciteSetBstMidEndSepPunct{\mcitedefaultmidpunct}
{\mcitedefaultendpunct}{\mcitedefaultseppunct}\relax
\EndOfBibitem
\bibitem[Ong \latin{et~al.}(2013)Ong, Richards, Jain, Hautier, Kocher, Cholia,
  Gunter, Chevrier, Persson, and Ceder]{Ong2013b}
Ong,~S.~P.; Richards,~W.~D.; Jain,~A.; Hautier,~G.; Kocher,~M.; Cholia,~S.;
  Gunter,~D.; Chevrier,~V.~L.; Persson,~K.~A.; Ceder,~G. Python Materials
  Genomics (pymatgen): A robust, open-source python library for materials
  analysis. \emph{Comput. Mater. Sci.} \textbf{2013}, \emph{68}, 314--319\relax
\mciteBstWouldAddEndPuncttrue
\mciteSetBstMidEndSepPunct{\mcitedefaultmidpunct}
{\mcitedefaultendpunct}{\mcitedefaultseppunct}\relax
\EndOfBibitem
\end{mcitethebibliography}

\end{document}